\documentclass[twocolumn]{aastex631}

\newcommand{\mr}{\mathrm}
\newcommand{\NNf}{[\ion{N}{2}]}
\newcommand{\NN}{[\ion{N}{2}] }
\newcommand{\OOf}{[\ion{O}{3}]}
\newcommand{\OO}{[\ion{O}{3}] }
\usepackage{soul}

\begin{document}

\title{The Fate of the Interstellar Medium in Early-type Galaxies. V. AGN Feedback from Optical Spectral Classification}

\correspondingauthor{Oleh Ryzhov}
\email{ryzhovuniv27@gmail.com}

\author{Oleh Ryzhov}
\affiliation{Astronomical Observatory Institute, Faculty of Physics and Astronomy, Adam Mickiewicz University, ul. S{\l}oneczna 36, 60-286, Pozna{\'n}, Poland}

\author{Micha{\l} J. Micha{\l}owski}
\affiliation{Astronomical Observatory Institute, Faculty of Physics and Astronomy, Adam Mickiewicz University, ul. S{\l}oneczna 36, 60-286, Pozna{\'n}, Poland}
\affiliation{Institute for Astronomy, University of Edinburgh, Royal Observatory, Blackford Hill, Edinburgh, EH9 3HJ, UK}

\author{J.~Nadolny}
\affiliation{Astronomical Observatory Institute, Faculty of Physics and Astronomy, Adam Mickiewicz University, ul. S{\l}oneczna 36, 60-286, Pozna{\'n}, Poland}
\affiliation{Instituto de Astrof\'isica de Canarias, E-38205 La Laguna, Tenerife, Spain}

\author{J.~Hjorth}
\affiliation{DARK, Niels Bohr Institute, University of Copenhagen, Jagtvej 155A, DK-2200 Copenhagen N, Denmark}

\author{A.~Le\'sniewska}
\affiliation{Astronomical Observatory Institute, Faculty of Physics and Astronomy, Adam Mickiewicz University, ul. S{\l}oneczna 36, 60-286, Pozna{\'n}, Poland}
\affiliation{DARK, Niels Bohr Institute, University of Copenhagen, Jagtvej 155A, DK-2200 Copenhagen N, Denmark}

\author{M.~Solar}
\affiliation{Astronomical Observatory Institute, Faculty of Physics and Astronomy, Adam Mickiewicz University, ul. S{\l}oneczna 36, 60-286, Pozna{\'n}, Poland}

\author{P.~Nowaczyk}
\affiliation{Astronomical Observatory Institute, Faculty of Physics and Astronomy, Adam Mickiewicz University, ul. S{\l}oneczna 36, 60-286, Pozna{\'n}, Poland}

\author{C.~Gall}
\affiliation{DARK, Niels Bohr Institute, University of Copenhagen, Jagtvej 155A, DK-2200 Copenhagen N, Denmark}

\author{T.~T.~Takeuchi}
\affiliation{Division of Particle and Astrophysical Science, Nagoya University, Furo-Cho, Chikusa-ku, Nagoya 464-8602, Japan}
\affiliation{The Research Center for Statistical Machine Learning, the Institute of Statistical Mathematics, 10--3 Midori-cho, Tachikawa, Tokyo 190-8562, Japan}

\received{14 August, 2024}
\revised{24 September, 2024}
\accepted{14 November, 2024}

\submitjournal{ApJS}

\begin{abstract}

Quenching of star-formation plays a fundamental role in galaxy evolution. This process occurs due to the removal of the cold interstellar medium (ISM) or stabilization against collapse, so that gas cannot be used in the formation of new stars. In this paper, we study the effect of different mechanisms of ISM removal. In particular, we revised the well-known Baldwin-Philips-Terlevich (BPT) and $\mr{EW_{H\alpha}}$ vs. \NNf$\mr{/H\alpha}$ (WHAN) emission-line ratio diagnostics, so that we could classify all galaxies, even those not detected at some emission lines, introducing several new spectral classes. We use spectroscopic data and several physical parameters of 2409 dusty early-type galaxies in order to find out the dominant ionization source [active galactic nuclei (AGNs), young massive stars, hot low-mass evolved stars (HOLMES)] and its effect on the ISM. We find that strong AGNs can play a significant role in the ISM removal process only for galaxies with ages lower than $10^{9.4}$\,yr, but we cannot rule out the influence of weak AGNs at any age. For older galaxies, HOLMES/planetary nebulae contribute significantly to the ISM removal process. Additionally, we provide the BPT and WHAN classifications not only for the selected sample but also for all 300\,000 galaxies in the GAMA fields.

\end{abstract}


\section{Introduction} \label{sec:intro}

Quenching of a galaxy, i.e.~when the star formation process significantly decreases or even stops, is an essential phase in the evolution of galaxies. The main reason for the galaxy becoming passive is the lack of interstellar medium (ISM) under proper conditions to form new stars \citep{Schawinski2014, Saintonge2022}.

ISM removal might be caused by several mechanisms. The
ISM can get incorporated in newly formed stars \citep[astration
;][]{Schawinski2014,Peng2015}, or it might be ionized or radiatively accelerated by 
supernovae \citep[SNe;][]{Dekel1986, Muratov2015}. Additionally, the ionization and heating by hot low-mass evolved stars \citep[HOLMES;][]{Stasinska_2008, Cid_Fernandes_2010, Herpich_2018} 
may remove the cold ISM
\citep{Herpich_2018}. Active galactic nuclei (AGN) feedback might act on the ISM by heating, ionizing, and expelling it \citep{Fabian_2012, Piotrowska2022}. 
The environment of a galaxy might also affect the ISM through, e.g.~merger events, heating processes in cluster environments, and ram pressure stripping \citep{Jachym2012, Mok2016, Sazonova2021}. 

Magnetic fields and turbulence  might also
have a similar effect
\citep{Padoan2002, Federrath2012}. Bulge of the galaxy might also prevent molecular gas from collapsing, stopping star formation  \citep[morphological quenching;][]{Martig2009,Gensior2020}.

Due to significant gas and dust production, destruction, and consumption in actively star-forming galaxies, it is challenging to explore their ISM removal process and the impact of each of the mechanisms described above. On the other hand, early-type galaxies (ETGs) are the most 
useful sample in this context,
due to their low SFRs 
\citep{M19}.

AGNs, young massive stars, HOLMES, and SNe 
leave distinctive features in the emission of the galaxy. One of the methods to determine the dominant ionization source in a galaxy 
is the application of emission-line diagrams, e.g.~\OOf$\mr{\lambda \lambda 5007/H\beta}$ vs.~\NNf$\mr{\lambda \lambda 6584/H\alpha}$, known as the Baldwin-Phillips-Telrevich diagram \citep[BPT;][]{bpt}; or equivalent width of $\mr{H\alpha}$
($\mr{EW_{H\alpha}}$) vs. \NNf$\mr{/H\alpha}$, known as the WHAN diagram \citep[WHAN;][]{Cid_Fernandes_2010, Cid_Fernandes_2011}.

The BPT diagram is based on flux ratios of commonly available emission lines. 
Its objective
is to separate galaxies with emission similar to that of star-forming regions with young massive stars \citep{Kewley2001, Stasinska2006, Sanchez2015}, and with emission originating from the AGN feedback \citep{Kauffmann2003, Stasinska2006}. Further semi-empirical analysis of galaxy emission resulted in the lower boundary for AGN hosts \citep{Kauffmann2003} and the upper limit for star-forming galaxies \citep{Kewley2001}. 
The classical BPT diagram is the most widespread and the most reliable emission-line diagnostic \citep{Stasinska2006}. 


The misclassification of AGN is a well-known problem of the BPT diagram. 
This includes
low-ionization nuclear emission-line regions \citep[LINER;][]{Heckman1980}, which tend not to have characteristic features of AGNs, except line ratios. The discussion on the origin of LINERs is still open: either they are caused by old stellar populations \citep{Stasinska_2008}, or by the specific type of AGNs \citep{Ho1999}. Additionally, according to the study of \citet{Agostino2019}, misclassification may occur for many other reasons, including low-ionization AGN, heavy dust attenuation, broadened Balmer series due to the broad-line region (BLR) emission, which could cause the galaxy to fall below the upper limit for SFGs at the BPT diagram, or  
so-called aperture effects \citep{Veilleux1995, Maragkoudakis2014, IglesiasParamo2016, Wylezalek2022}. However, the most recent results of \citet{Agostino2023_SDSS} suggested a high correlation ($\sim80-90\%$) between X-ray AGNs and AGNs that were classified using emission lines. 

Aperture effects might occur for single-aperture spectroscopic surveys that cover a considerable part of a galaxy, combining different types of emission in different regions. Thus, AGN emission might be diluted by star-forming regions \citep{Agostino2019} or by diffused ionized gas \citep[DIG;][]{Wylezalek2022}. 

The WHAN diagram is a more recent method of spectral classification of galaxies. Instead of four emission lines, which should be detected 
\citep{Kewley2006}, the WHAN relies on two of the most common ones: \NN 
and $\mr{H\alpha}$. This method allows classifying two times more galaxies, including those that have non-detected \OO and/or $\mr{H\beta}$ lines \citep{Cid_Fernandes_2011}. In addition to that, the WHAN also provides information on so-called retired galaxies, i.e.~galaxies with quenched star-formation that tend to be ionized by HOLMES.

This is the fifth paper in a series that is dedicated to studying the ISM removal mechanism in early-type (i.e.~elliptical, lenticular, and early-type spiral) galaxies. In the first paper \citep{M19}, we presented the dust-to-stellar mass decline, as a function of light-weighted stellar age, and measured the dust removal timescale [$\tau = (2.5 \pm 0.4)$ Gyr]. In \citet{Lesniewska2023} we expanded the studied sample to $\sim 2000$ ETGs with detected dust. Using that sample, we analyzed what galaxy properties affect the dust removal process, and estimated the dust removal timescale to be $\tau = (2.26 \pm 0.18)$ Gyr, which was in agreement with the results from the previous work. Next, in \citet{Michalowski2024} we made a major exploration of the effectiveness of different ISM removal mechanisms and sources of energy powering these mechanisms, based on 
CO 
and \ion{H}{1} observations. That resulted in reducing the list of viable ISM removal mechanisms in ETGs to the ionization and heating by AGNs, supernovae type Ia, and hot low-mass stars. In \citet{Nadolny2024}, we find that simulated galaxies have similar cold ISM removal timescales to the observed ones, with SN type II feedback dominating the removal at earlier times, while later astration and cold gas heating by stellar feedback becoming dominant. The goal of this paper is to determine the dominant mechanism of the ISM decline in ETGs using spectral analysis.

The structure of this paper is the following. Section \ref{sec:data} presents the ETG sample, Section \ref{sec:method} includes the description of the spectral classification we implemented. Results are presented in Section \ref{sec:res}. The discussion of AGN and old stellar population feedback is presented in Section \ref{sec:disc}, together with statistical biases that could affect spectral classification. We close with a summary in Section \ref{sec:conc}.




\section{Data and Sample} \label{sec:data}

The Galaxy And Mass Assembly survey\footnote{\url{http://www.gama-survey.org}} (GAMA; \citealt{Driver2011,Driver2016,Baldry2018,Smith2011}) is aimed at studying cosmology and galaxy formation and evolution by conducting observations with the latest generation of telescopes, both ground-based and space-based. This survey provided optical spectroscopic observations with the AAOmega multi-object spectrograph at the Anglo-Australian Telescope (AAT) of $\sim$300\,000 galaxies $(r < 19.8^m)$. 

We selected dusty early-type galaxies from the GAMA survey Data Release 4 \citep[DR4;][]{Driver2022} using the criteria, described in \citet{Lesniewska2023}: Sérsic index $n > 4$ \citep{Sersic1963, Kelvin2012}, signal-to-noise at the \textit{Herschel} SPIRE \citep{spire} 250 $\mu$m band greater than 3, and redshift $0.01 < z < 0.32$.

We made use of optical emission line properties of these galaxies, i.e.~fluxes and equivalent widths of $\mr{H{\alpha}}$, $\mr{H{\beta}}$, \OOf, \NN \citep{Gordon2017}\footnote{\url{www.gama-survey.org/dr3/data/cat/SpecLineSFR/}}.
 
For our analysis, we used spectra obtained from GAMA or SDSS because they were flux-calibrated, thus we did not analyze the data for 38 galaxies observed by 2dFGRS, MGC and 6dFGS. Moreover, 4 galaxies observed by the SDSS survey had technical problems with extracting line properties ($-99999.0$ flag for either equivalent width or line flux). Finally, we ended up with 2409 galaxies.

The GAMA catalog also contains estimation of several physical properties of galaxies and their uncertainties. In particular, for this work we make use of $r$-band light-weighted stellar age (hereafter age), stellar mass ($M_{\rm stellar}$), star formation rate (SFR) averaged over the last $100 \, {\rm Myr}$, and cold and warm dust temperatures. All these parameters rely on the analysis of broad-band photometry data and spectral energy distribution using the MAGPHYS software \citep{Cuhna2008}, hence are independent of emission lines.

Following \citet{Lesniewska2023}, we also distinguished galaxies on the main sequence (MS; as measured by \citealt{Speagle_2014}) and below it. This resulted in 1539 below-MS galaxies and 870 MS galaxies.

\section{Method}\label{sec:method}




\subsection{BPT Diagnostics}\label{BPTALGO}

\begin{figure*}
\centering
\includegraphics[width=1.0\textwidth]{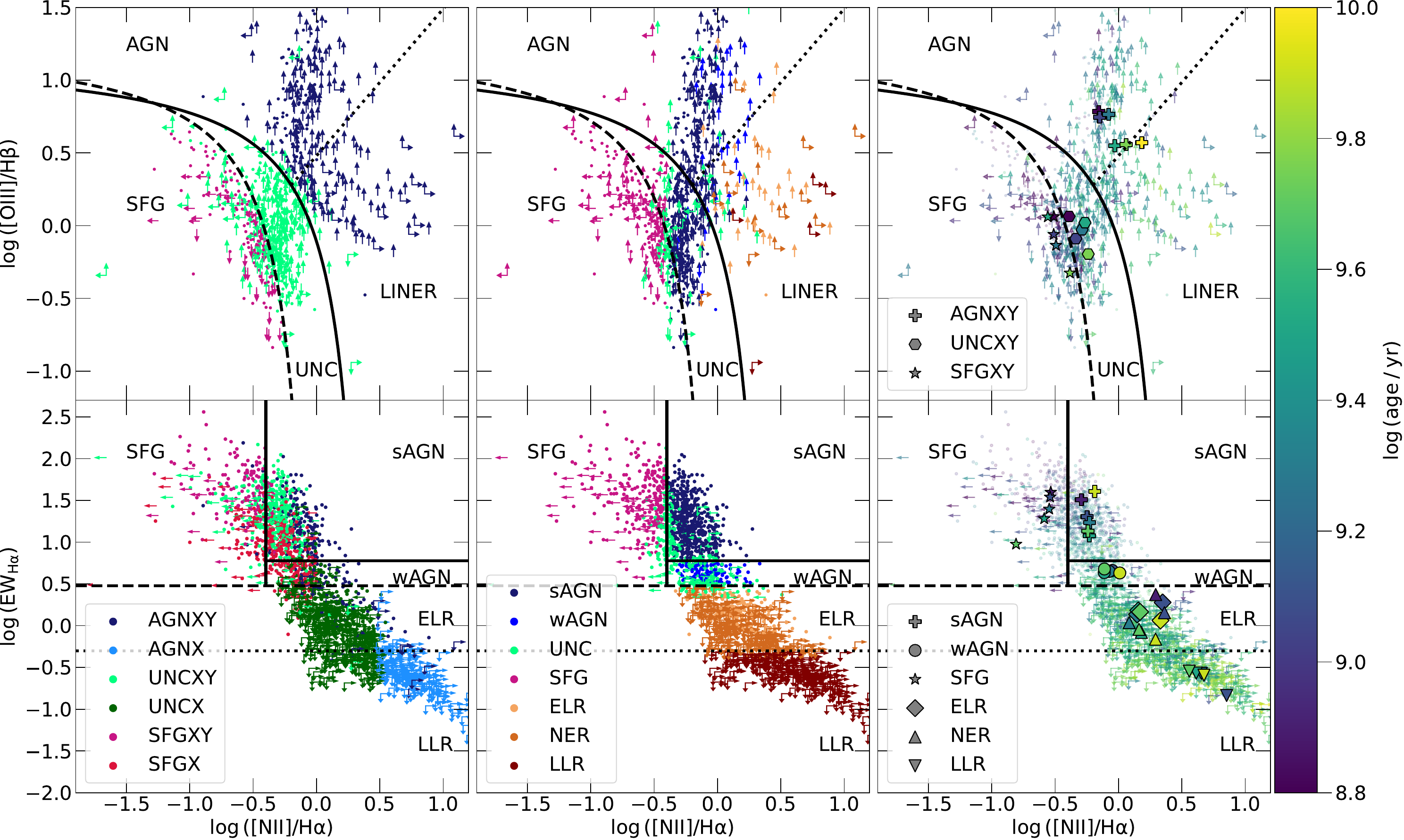}
\caption{BPT (\citealt{bpt}, top panels) and WHAN (\citealt{Cid_Fernandes_2011}, bottom panels) emission-line ratio diagnostic diagrams for the studied sample. The BPT diagrams contain the lower limit for active galactic nuclei hosts (\citealt{Kauffmann2003}, black solid line), upper limit for star-forming galaxies (\citealt{Kewley2001}, black dashed line), and separation line between Seyfert objects and LI(N)ERs (\citealt{Cid_Fernandes_2010}, black dotted line). The WHAN diagram contains delimitation lines and spectral classes, described in \citet{Herpich_2016}.
The left panels show the sample color-coded by BPT spectral classes (Section~\ref{BPTALGO}), whereas the middle panels present the sample color-coded by the WHAN spectral classes (Section~\ref{WHANALGO}). The panels on the right display the galaxies of the studied sample color-coded correspondingly to their stellar light-weighted age value. The median positions of the spectral classes in each age bin are also presented (see Section \ref{res:agesnap}). The information on colors and markers for spectral types is placed in the legends. Both BPT and WHAN are used to classify galaxies into active galactic nuclei (AGN) hosts, star-forming galaxies (SFG), and galaxies, which emission allows both (UNC). The -XY flag for the BPT class implies that it was identified using both ratios, whereas the -X flag that only the \NNf/${\rm H\alpha}$ ratio was used due to the non-detection of the other two lines. The WHAN diagram additionally provides different classes of retired galaxies (RGs): line-less retired (LLR), emission-lines retired (ELR), and noisy emission retired (NER)}
\label{fig:BPT_WHAN}
\end{figure*}

The top panels of Fig.~\ref{fig:BPT_WHAN} present the classical BPT diagrams for our galaxy sample.

The \citet{Kauffmann2003} line was used as the lower limit for AGN-dominated galaxies, whereas the \citet{Kewley2001} results were used in the selection of star-forming galaxies (SFG). They are one of the most widely used delimitation lines for the BPT diagram. The AGN/LI(N)ER classification method of \citet{Cid_Fernandes_2010} is shown only for the reference. We do not use this delimitation line in our analysis due to the numerous lower limits for the objects located in the LINER area. In other words, we are not able to state whether the emission of these objects is similar to LINER or AGN.

We treated lines as detected if 
the signal-to-noise (SN) ratio is at least $2$. Setting this value to $3$ neither changes the results significantly nor affects our interpretation and conclusions.

Besides the usual cases with the detection of all four emission lines, there are four special cases in our classification. First, if one of the lines entering the ratio for a given axis is not detected, then we use a $\mr{2\sigma}$ upper or lower limit for the ratio. Second, if a given line is an absorption line, then we treat it as a non-detection using the reported flux errors. Third, if both lines entering the ratio for a given axis are not detected (or are absorption lines), then we only use the information from the other axis (see the 1D cases below). Fourth, if neither of the four lines is detected, then we classify the galaxy as 'no emission lines' (NOEL). 

From all the possible cases described above, we get the following spectral classes based on the BPT diagram:

\begin{itemize}
    \item Star-forming galaxies (SFGs).
    \begin{itemize}
        \item SFGXY. XY denotes using both the x- and y-axes of the diagnostic diagram to place the galaxy inside the SFG region,  
        defined by \citet{Kewley2001}, either within their error bars or $2\sigma$ limits. This is the case for {\bf 157 galaxies}.
        \item SFGX. These galaxies are undetected at \OO and $\mr{H\beta}$, so we can only use the x-axis information and for them the ratio $\log($\NNf$\mr{/H\alpha}) < 0.05$. This limit is the asymptotic value of the SFG curve of  \citet{Kewley2001}. In this category, the $\mr{H\alpha}$ line needs to be detected (otherwise the x-axis ratio would be a lower limit), but the \NN line does not. 
        There are {\bf 444 galaxies} in this category.
        \item SFGY. These galaxies are undetected at $\mr{H\alpha}$ and \NN lines, so only the y-axis information can be used. If the y-axis ratio is $\log($\OOf$/\mr{H\beta}) < 1.3$ (the asymptotic value of the curve of \citealt{Kewley2001}), this places them in the SFG region. To satisfy this, $\mr{H\beta}$ line needs to be detected, but \OO does not. There are {\bf 5 galaxies} in this category.
    \end{itemize}
    \item Active Galactic Nuclei.
        \begin{itemize}
        
        \item AGNXY. As in the case of SFGXY, both x- and y-axis information is used, but in this case, this places them above the \citet{Kauffmann2003} line, considering either the ratio limits or error bars. This happens for {\bf 262 galaxies}.
        \item AGNX. As SFGX, these galaxies are undetected at \OO and $\mr{H\beta}$, so we only use the x-axis information and they have $\log($\NNf$\mr{/H\alpha}) > 0.47$ (the asymptotic value of the \citealt{Kauffmann2003} line). \NN lines needs to be detected, but the $\mr{H\alpha}$ does not. There are {\bf 160 galaxies} in this category.
        \item AGNY. As SFGY, these galaxies are undetected at \NN and $\mr{H\alpha}$, but they should have $\log($\OOf$\mr{/H\beta}) > 1.19$ (the asymptotic value of the \citealt{Kauffmann2003} line). \OO line needs to be detected, but $\mr{H\beta}$ does not. There are no galaxies in this category.

        \end{itemize}

    \item Unclear cases (UNC), for which data allow both SFG and AGN classifications. This includes galaxies with error bars spanning both SFG and AGN regions, or with a ratio lower limit below the AGN line, or with a ratio upper limit above the SFG line. If both x- and y-axis information is available, they are denoted UNCXY ({\bf 386 galaxies}). If only x- or y-axis information is used, then they are denoted UNCX ({\bf 470 galaxies}) and UNCY ({\bf 11 galaxies}), respectively.  

    \item NOEL (no emission lines galaxies). These are the galaxies that are undetected at \OOf, \NNf, $\mr{H\alpha}$, and $\mr{H\beta}$ lines. This happens for {\bf 514 galaxies}.
    
\end{itemize}

Appendix~\ref{ax:corr} contains the analysis of the nature of each presented spectral class, which revealed that AGNX and UNCX appeared to be passive galaxies with a lack of 2-3 emission lines, while AGNXY and UNCXY show signs of ongoing ionization processes.

\subsection{WHAN Diagnostics}\label{WHANALGO}

The bottom panels of Fig.~\ref{fig:BPT_WHAN} display the WHAN diagram. The classification and the dividing lines include strong/weak AGNs, SFGs, and emission-line retired and line-less retired galaxies \citep{Cid_Fernandes_2011, Herpich_2016}. Similarly to our classification using the BPT diagram, here we used the upper and lower limits,  adding the classes of noisy equivalent width retired galaxies (not being able to distinguish emission-line and line-less options) and unclear cases. 
Thus, we ended with the following classes:

\begin{itemize} 
    \item sAGN: strong AGN with $\log($\NNf$\mr{/H\alpha) > -0.4}$ and $\mr{EW_{H\alpha} > 6 \AA}$. ({\bf 553 galaxies}).
    \item wAGN: weak AGN with $\log($\NNf$\mr{/H\alpha) > -0.4}$ and $\mr{3\AA < EW_{H\alpha} < 6 \AA}$. ({\bf 123 galaxies}).
    \item SFG: star-forming galaxies with $\log($\NNf$\mr{/H\alpha) < -0.4}$ and $\mr{EW_{H\alpha} > 3 \AA}$. ({\bf 257 galaxies}).
    \item UNC: unclear cases. This includes galaxies above the wAGN line $\mr{(EW_{H\alpha} > 3 \AA)}$, and either an upper limit for \NNf$\mr{/H\alpha}$ within the AGN regime, or a lower limit in the SFG regime, or an upper limit for $\mr{EW_{H\alpha}}$. Additionally, galaxies with $\mr{1\sigma}$ uncertainties crossing the delimitation lines were also included in this category. This is the case for {\bf 304 galaxies}.
    \item Retired galaxies (RGs):
    \begin{itemize}
        
    \item ELR: emission-line retired galaxies with $\mr{0.5 \AA < EW_{H\alpha} < 3 \AA}$ ({\bf 194 galaxies}).
    \item LLR: line-less retired galaxies with $\mr{EW_{H\alpha} < 0.5 \AA}$ ({\bf 388 galaxies}).
    \item NER: noisy equivalent width retired galaxies. The galaxy is assigned to this class if it is retired, but it is unclear if it is ELR or LLR, i.e.~if it has an upper limit for $\mr{EW_{H\alpha}}$, which lies in between the LLR limit and wAGN limit $\mr{(0.5 \AA < EW_{H\alpha} < 3 \AA)}$. This category includes {\bf 590 galaxies}. 
    
    \end{itemize}
\end{itemize}

\section{Results} \label{sec:res}

\begin{figure*}
\centering
\includegraphics[width=0.95\textwidth]{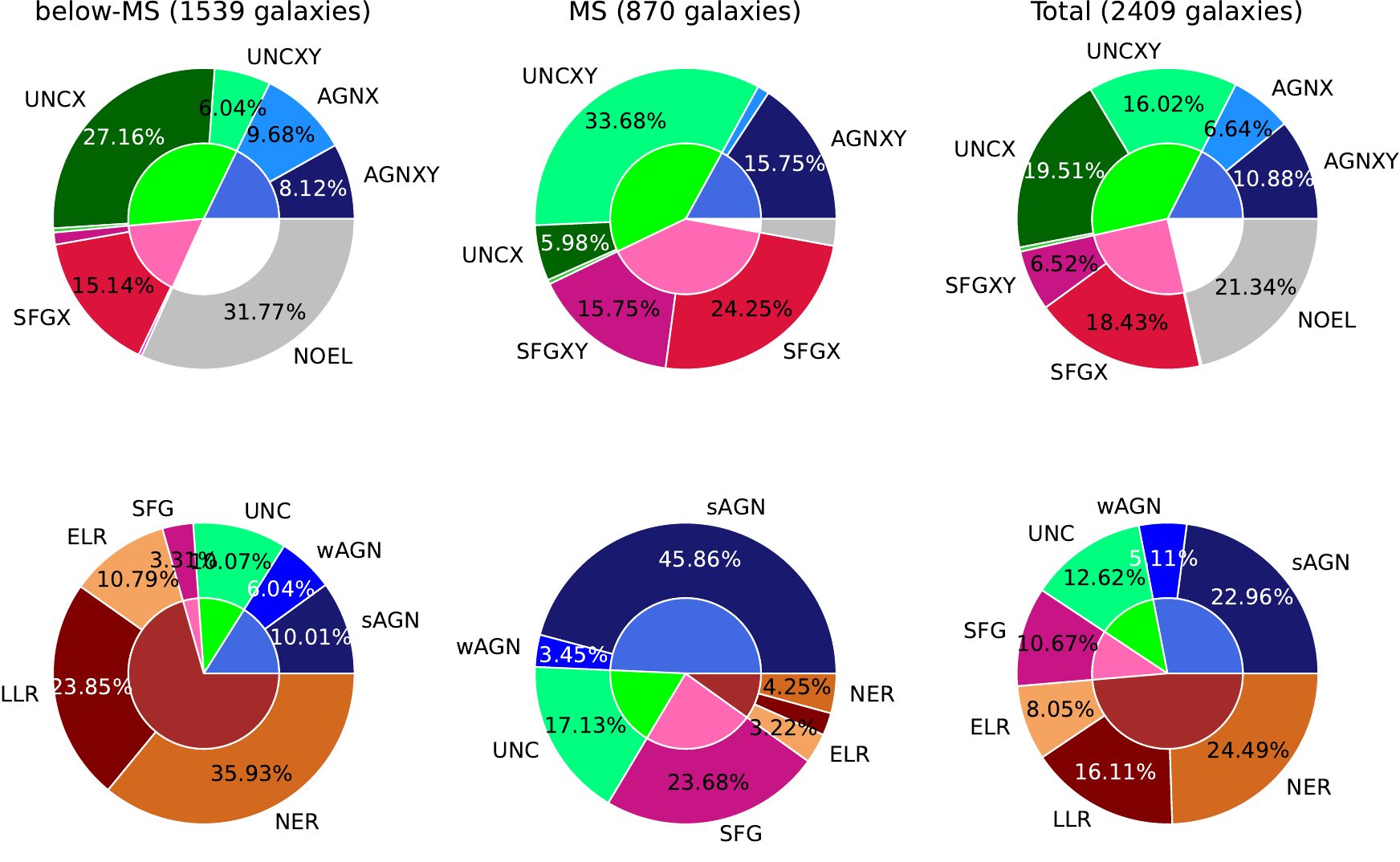}
\caption{Distribution of BPT (top panels) and WHAN spectral classes (bottom panels) in different studied subsamples: below-MS galaxies (left panels), MS galaxies (middle panels), and total sample (right panels). Outer circles contain detailed information on spectral classes, whereas inner circles show the summarized statistics for AGNs (blue), SFGs (pink), unclear cases (green), and NOEL/RGs (grey/brown).
\label{fig:BMSMS}}
\end{figure*}

\subsection{Spectral Classification}
\label{res:distr}

In this section, we show the general distribution of spectral classes that were described above for dusty ETGs. Moreover, we explore the differences between the below-MS and MS galaxies. 

Figure~\ref{fig:BMSMS} presents how our galaxies are classified via the BPT diagram (top panels) and the WHAN diagram (bottom panels). The left panels contain information about 1539 below-MS galaxies, the middle panels about 870 MS galaxies, and the right panels present the distribution of the whole sample.

The BPT diagnostic resulted in $21\%$ of NOEL, $18\%$ of SFGX, and $7\%$ of SFGXY ($25\%$ of SFG galaxies in total). The fraction of UNCX ($20\%$) and UNCXY ($16\%$) make UNC galaxies account for $36\%$ of the sample. AGNs in total occupy only $18 \%$ of the sample, among that only $11\%$ are AGNXY. The galaxies that were classified only by the $\log($\OOf$\mr{/H\beta})$ ratio (`Y' class) do not constitute a significant fraction ($<1\%$) of our sample.

The WHAN classes show slightly different distributions: $23\%$ of sAGN, $5\%$ of wAGN, $11\%$ of SFG, $8\%$ of ELR, $16\%$ of LLR, $25 \%$ of NER ($49\%$ of RG in total), and, finally, $13\%$ of UNC.


The BPT classification revealed that the MS subsample, as expected, contains very few NOEL galaxies ($<3\%$), while the below-MS subsample has $32\%$ of them. Moreover, a big difference exists between the ratio of the fraction of the XY classes and the X-axis classes: e.g.~there are $34\%$ and $6\%$ UNCXY and UNCX galaxies in the MS and $6\%$ and $27\%$ in the below-MS sample, respectively. The same can be observed for AGN and SFG classes. This is expected given that MS galaxies should have more line detections.

WHAN classes behave in an even clearer way: the MS subsample 
is dominated by SFG, wAGN, and sAGN ($\sim 75\%$), with only $10\%$ of retired galaxies, while the below-MS subsample is mostly composed of retired galaxies, leaving only $\sim30\%$ to SFG, UNC, wAGN, and sAGN. In addition, the ratio of the numbers of sAGN to wAGN is different for the below-MS and MS samples: the MS sample contains 13 times more sAGNs than wAGNs, while in the below-MS sample there are $\sim 2$ times more sAGNs, than wAGNs.

\subsection{BPT and WHAN Classes in Age Bins}
\label{res:agesnap}

\begin{figure*}[p!]
\centering
\includegraphics[width=0.8\textwidth]{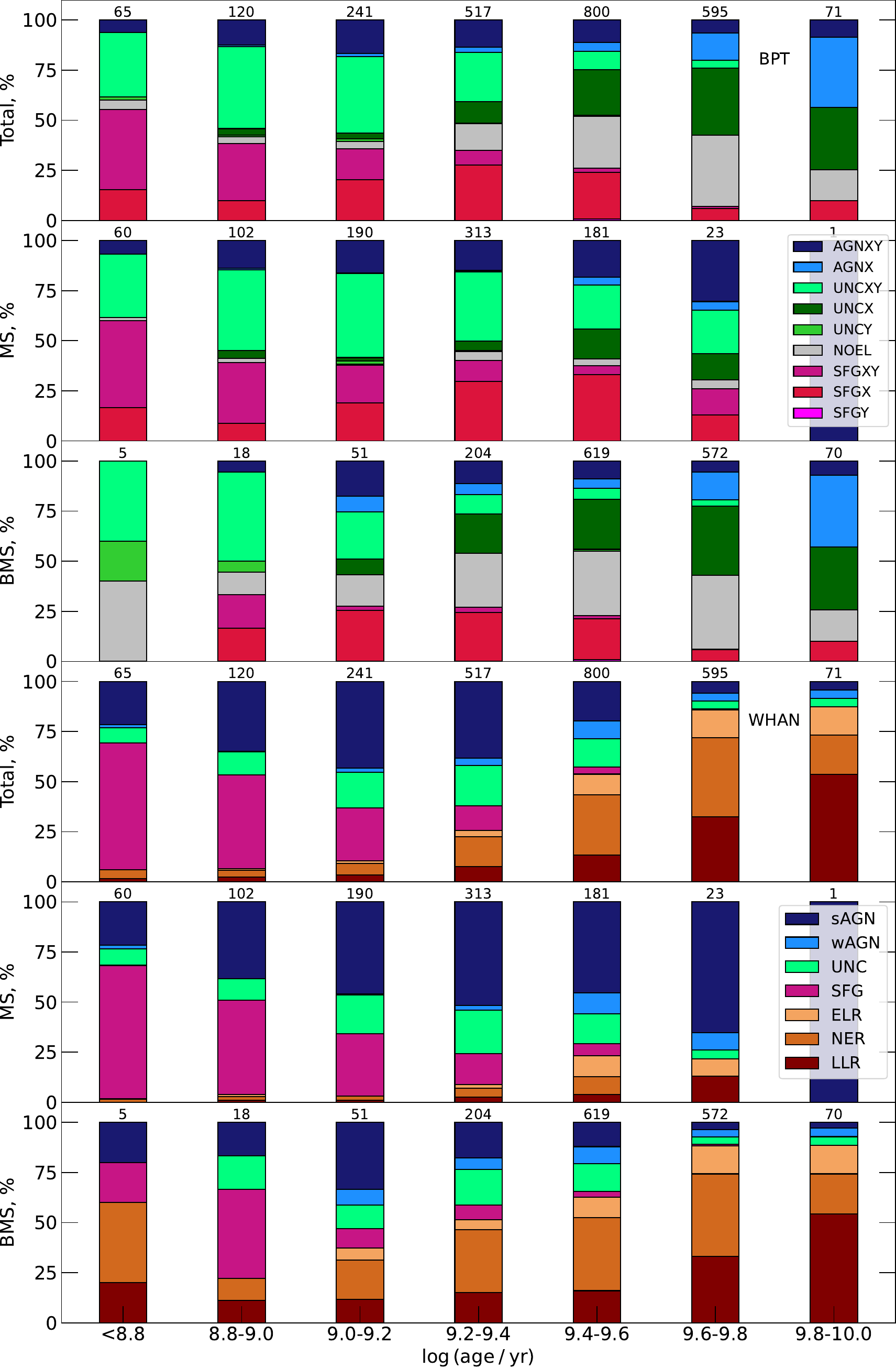}
\caption{BPT (top three panels) and the WHAN (bottom three panels) spectral class distribution as a function of the light-weighted age. In each group of panels, the top panel presents the distribution of the whole studied sample, the middle corresponds to the MS subsample, and the bottom displays the below-MS subsample.}
\label{fig:AGE_BINS}
\end{figure*}

We analyze systematic changes in the fraction of each spectral class as a function of light-weighted stellar age. The sample was divided into seven age bins: $\log({\rm age}/{\rm yr})<8.8$ and six bins with widths of $0.2$\,dex, 
up to $10.0$.
The analysis was performed for the whole sample, MS, and below-MS subsamples. 

Figure~\ref{fig:AGE_BINS} presents the fraction of spectral classes as a function of age. Taking into account the BPT classes for the whole sample (top panel), it is evident that the fraction of SFGs steadily decreases with age, as expected. 
The fraction of UNCXY drops with age, while the percentage of UNCX galaxies generally increases, making the total UNC category relatively constant at $30\%$.
AGNXY and AGNX have different behavior. AGNXY, starting from $\log({\rm age}/{\rm yr})=9.2-9.4$ shows a drop from $17\%$ to $9\%$, while AGNX increases from $2\%$ to $35\%$. Such a difference between AGNXY and AGNX might be explained by the different nature of the ongoing ionization processes in them (see Appendix~\ref{ax:corr}).

Finally, the fraction of NOEL galaxies increases with age, peaking at the level of $35\%$ in the $9.6-9.8$ age bin, which is a well-known fact explained by the lack of cold gas in older, passive galaxies (e.g.~\citealt{Spilker2018, Michalowski2024}).

The WHAN classification of the whole sample (fourth panel from the top in Fig.~\ref{fig:AGE_BINS}) shows the same trend of a decreasing fraction of SFGs with age. All types of RGs (LLR, ELR, NER) show a drastic growth from $5\%$ in the $8.8 - 9.0$ $\rm{dex(yr)}$ age bin to $90\%$ in the last bin. 
The AGN classes (sAGN and wAGN) grow slightly in the first bins, reaching $\sim40\%$ for sAGN in the $9.2-9.4$ $\rm{dex(yr)}$ bin and $\sim10\%$ for wAGN in $9.4-9.6$ $\rm{dex(yr)}$, and then drop down to $4\%$ each in the last bin. 

The below-MS sample exhibits the decline of SFGs for increasing age for both the BPT and WHAN classifications, as expected. This is compensated by an increase of NOEL, AGNX, and UNCX galaxies in the BPT diagram and retired galaxies in the WHAN diagram.
The MS galaxies show a similar decline of SFGs, but in this case this is compensated by the rise of AGN classifications (AGNXY for BPT and sAGN for WHAN).

\subsection{Dust Mass and Temperatures}
\label{res:temp}

\begin{figure*}
\centering
\includegraphics[width=0.9\textwidth]{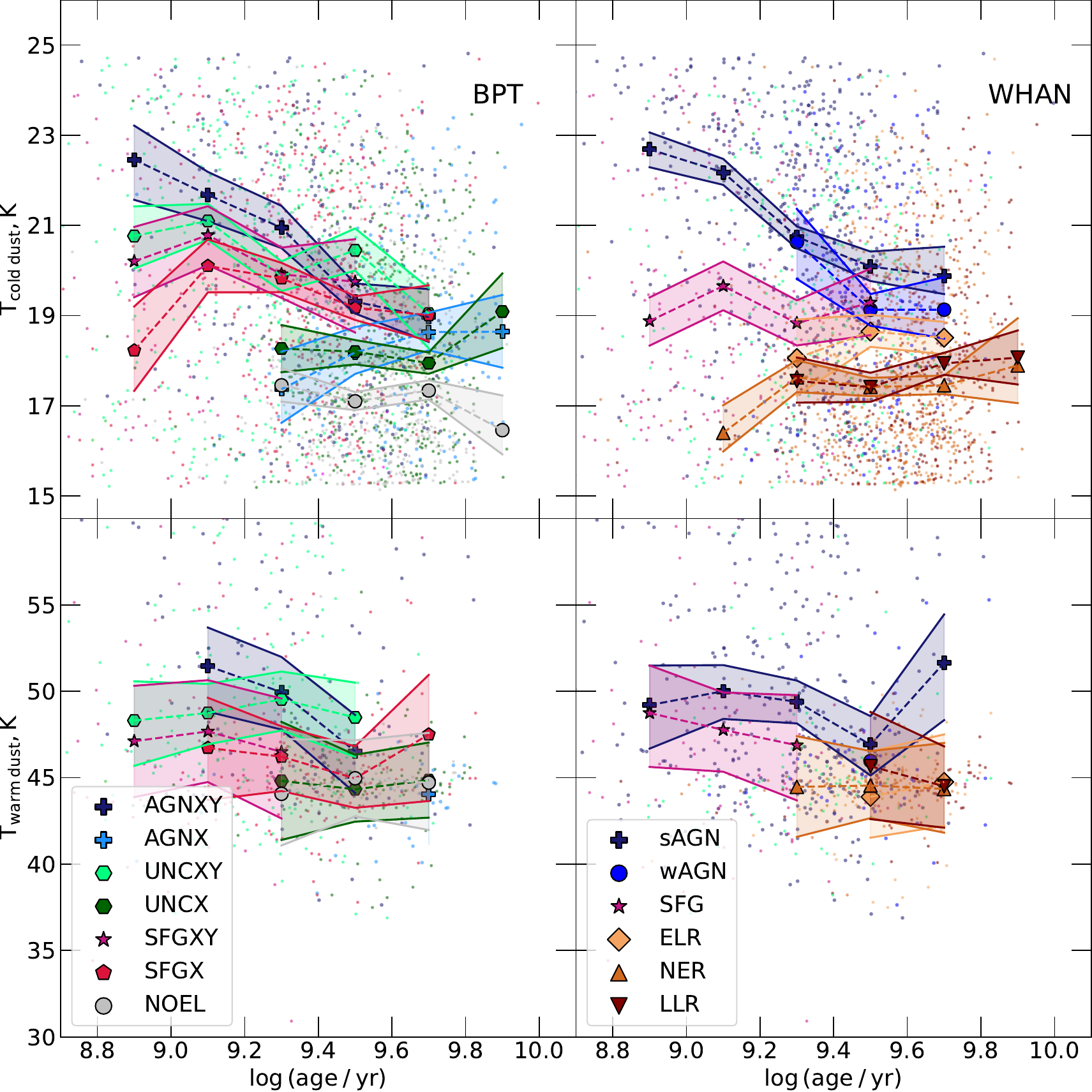}
\caption{Cold dust temperature (top panels) and warm dust temperature (bottom panels) as a function of light-weighted stellar age, considering BPT (left) and WHAN spectral classes (right). For the warm dust temperature, the galaxies with the lack of a detection in \textit{Herschel} PACS 100 ${\rm \mu m}$ were excluded. Big symbols present median values of temperature for galaxies of correspondent spectral class in each age bin (see Section \ref{res:temp}). In order to distinguish trends, the median values are connected using dashed lines. Derived uncertainties are displayed as color-filled areas around medians, limited by solid lines.
\label{fig:TEMP}}
\end{figure*}

\begin{figure*}
\centering
\includegraphics[width=0.95\textwidth]{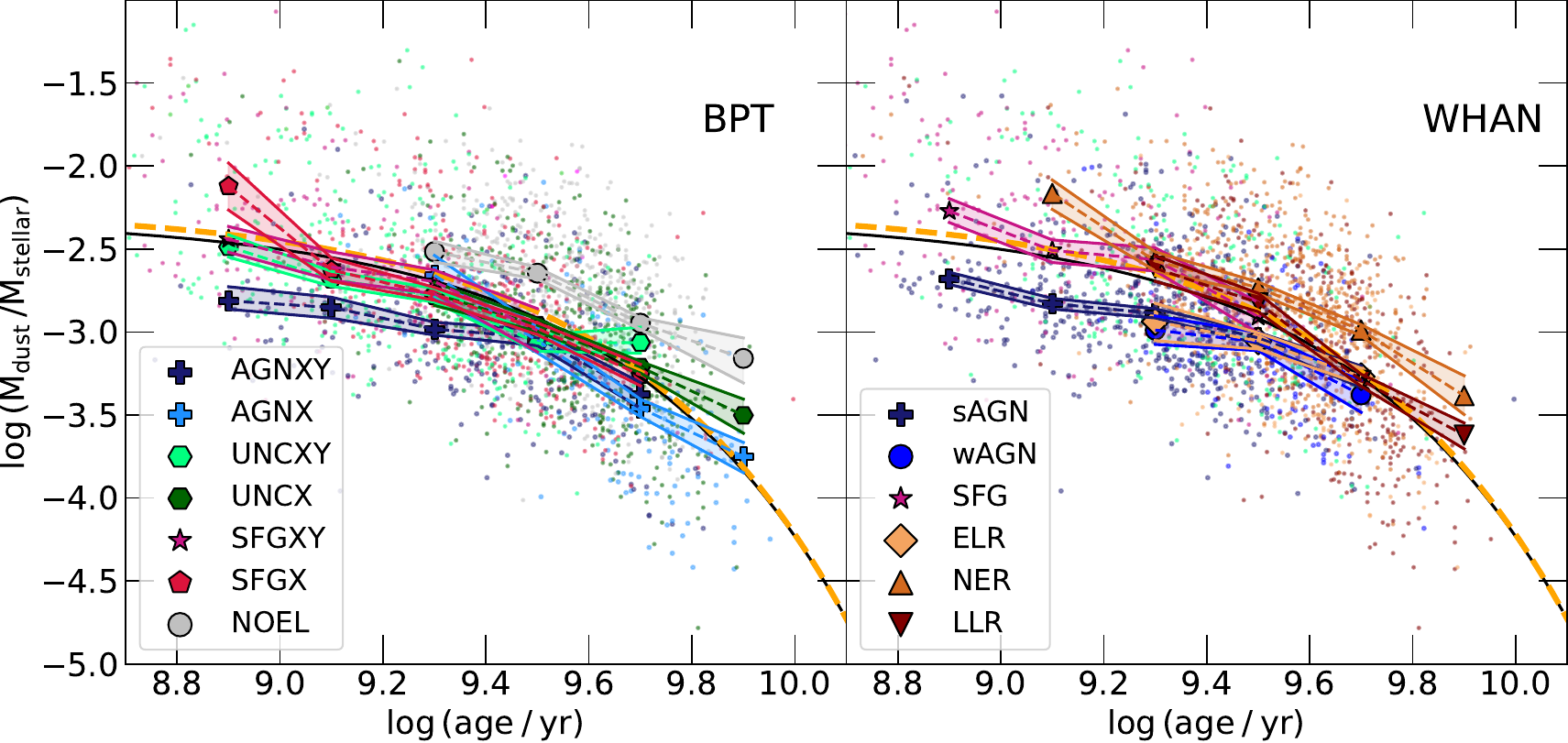}
\caption{The ratio of dust mass to stellar mass as a function of stellar age in terms of BPT (left) and WHAN classifications (right). The galaxies are color-coded correspondingly to their spectral classes, in addition,  median values for each class in each age bin are presented as the big symbols. Both figures show the dust decline function estimated by \citet{Lesniewska2023} (black solid line).
\label{fig:MDMS}}
\end{figure*}

Figure~\ref{fig:TEMP} shows the temperature of cold (top) and warm (bottom) dust from the MAGPHYS SED modeling vs.~light-weighted stellar age color-coded by the BPT (left panels) and WHAN classes (right panels). For better visualization of the trends, we calculated the median values of temperatures in previously defined age bins (Sec.~\ref{res:agesnap}) for all spectral classes, excluding 
the youngest bin. Galaxies with $\log{(\text{age/yr})} < 8.7$ are not presented in the figures due to their low number (42 in total). 

To estimate errors on the medians, we used the Monte-Carlo simulation to perturb data points of each spectral class in each age bin 1000 times by a value drawn from a Gaussian distribution with a width equal to the temperature uncertainty. We considered the 16 and 84 percentiles of the resulting distributions as uncertainties of the medians. The Monte-Carlo simulations were not performed for bins containing less than 10 galaxies of the correspondent spectral class.

Galaxies with S/N $<2$ in the \textit{Herschel} PACS 100 ${\rm \mu m}$ filter were not included in the warm dust temperature analysis because, without a detection at shorter infrared wavelengths, the warm dust content is almost unconstrained. For the studied sample, 733 galaxies are detected, other 111 were not observed, and 1565 were observed, but non-detected at the \textit{Herschel} PACS 100 ${\rm \mu m}$ filter.

The top left panel of Fig.~\ref{fig:TEMP} shows that the cold dust temperature decreases with age for BPT AGNXY, SFGXY, SFGX, and UNCXY classes. The temperatures of NOEL, AGNX, and UNCX galaxies are relatively constant with age and are lower than those of other classes. AGNXY are slightly hotter than SFGs only in the youngest bin. The cold dust temperature of UNCXY galaxies fluctuates at the level of the median temperatures of AGNXY and SFGXY.

The temperature of cold dust in WHAN sAGNs and BPT AGNs 
decreases significantly with age.
 
These AGNs exhibit significantly higher cold dust temperatures than SFGs, especially in younger age bins ($\log \rm{(age / yr)} < 9.4$). RGs generally do not change their cold dust temperature with age, however, we observe a systematic increase of temperature for LLRs for higher ages, but the trend is weak.

Much weaker trends can be seen for warm dust temperatures, which might be explained by the fact that only 733 galaxies ($30\%$ of the studied sample) are detected at PACS 100 ${\rm \micron}$. The warm dust temperatures of WHAN sAGNs and BPT AGN 
decrease slightly with age. 
The remaining classes
do not change warm dust temperatures with age significantly.

Figure~\ref{fig:MDMS} displays the dust removal phenomenon, i.e.~the decreasing ratio of dust-to-stellar mass with age for dusty ETGs \citep{M19, Michalowski2024,Lesniewska2023, Nadolny2024}. The figure also contains the 
median values and their uncertainties for each spectral class in every age bin, in order to better identify trends.

According to Fig.~\ref{fig:MDMS}, each spectral class possesses lower mass fractions of dust for higher ages. No clear differences between spectral classes are visible, but BPT NOEL and WHAN NERs possess slightly higher dust-to-stellar mass ratios in correspondent age bins compared to other spectral classes. Such a tendency might be explained by the fact, that the BPT NOEL and WHAN NERs are the coldest galaxies in the sample (Fig.~\ref{fig:TEMP}), and the faintest in the \textit{Herschel} SPIRE 250 ${\rm \mu m}$ filter. 
Thus, we select only those NOELs and NERs, that contain a lot of dust, due to the set threshold on the signal-to-noise ratio in the S250 for dusty ETGs (Section~\ref{sec:data}).

\subsection{SFR and Stellar Mass}
\label{res:sfrsm}

\begin{figure*}
\centering
\includegraphics[width=0.8\textwidth]{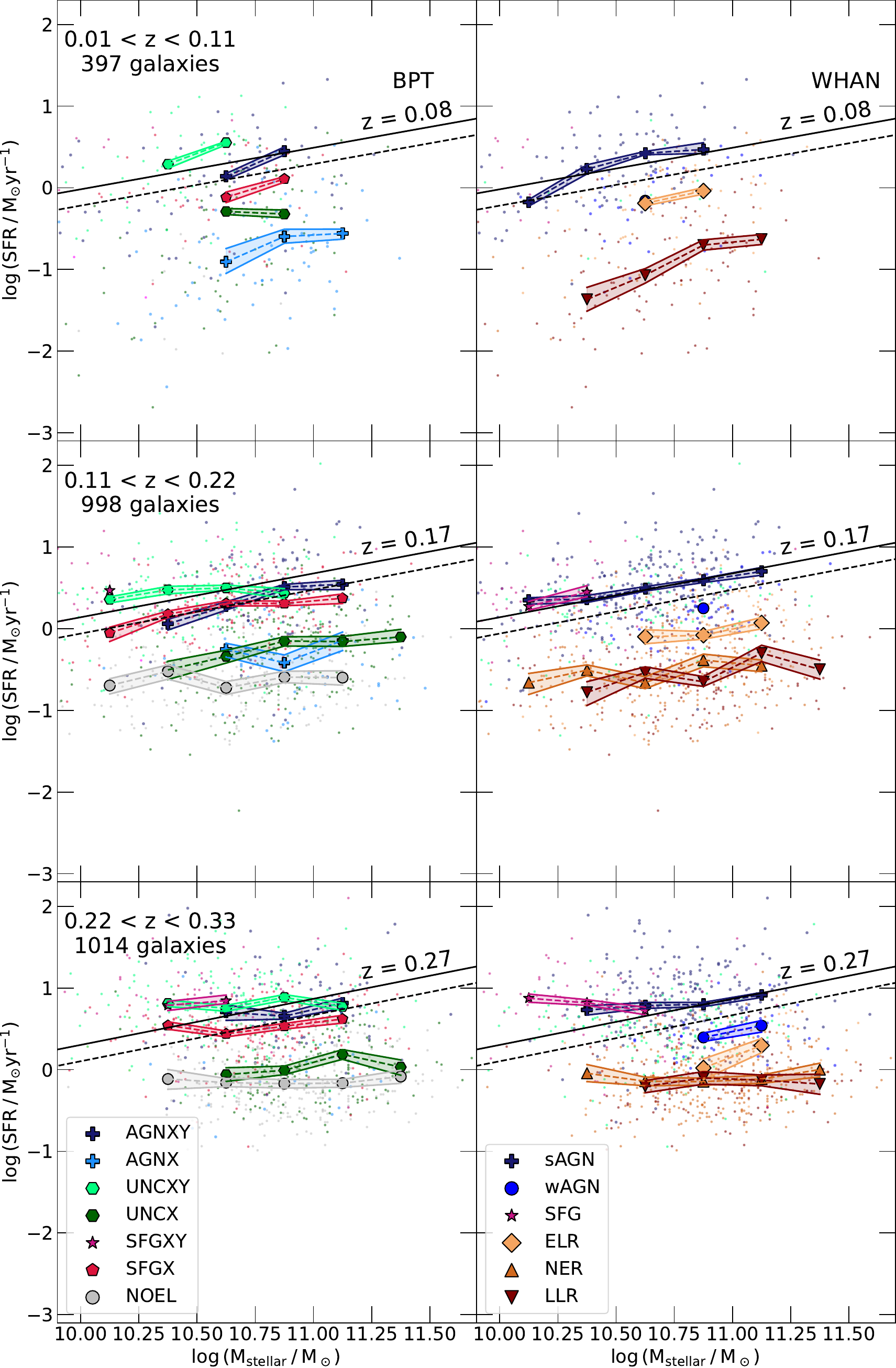}
\caption{Star-formation rate averaged over the last 100 Myr as a function of the stellar mass of a galaxy for the BPT (left column) and the WHAN (right column) spectral classes. The top panels display the galaxies with $0.01 < z < 0.11$, the middle with $0.11 < z < 0.22$, and the bottom with $0.22 < z < 0.33$. Black solid lines represent the Main Sequence \citep{Speagle_2014}, calculated, considering the median redshift value for the subsample. Black dashed lines demarcate the lower limit for MS-galaxies, shifted by $-0.2 \; {\rm dex}$ from the MS. Similarly to Fig.~\ref{fig:TEMP}, markers demonstrate median values of SFR for each spectral class in every mass bin (see Section \ref{res:sfrsm}). Uncertainties are presented as color-filled areas, limited by solid lines.
\label{fig:SFRSM}}
\end{figure*}

Figure~\ref{fig:SFRSM} presents SFR averaged over the last 100 Myr from the MAGPHYS SED modeling as a function of the stellar mass of a galaxy. 
The MS evolves with cosmic time, so we split the studied sample into three redshift ranges: $0.01 < z < 0.11$, $0.11 < z < 0.22$, and $0.22 < z < 0.33$. For better visualization, we additionally divided our sample within $\log \mathrm{(M_{stellar}/M_\odot)}=10.0$ to $11.5$ into 6 bins with the width of $0.25$\,dex, and calculated median values with errors, following the same method as described in Section \ref{res:temp}. 


As expected, SFGs and AGNs for both BPT and WHAN classifications are within the main sequence, while WHAN RGs and BPT classes lacking strong emission lines (NOEL, UNCX, AGNX), lie below the main sequence at all panels of Fig.~\ref{fig:SFRSM}. WHAN wAGN and BPT SFGX have SFRs in between these active and less-active galaxies. 

\subsection{BPT and WHAN Classes in Stellar Mass Bins}

\begin{figure*}[p!]
\centering
\includegraphics[width=0.8\textwidth]{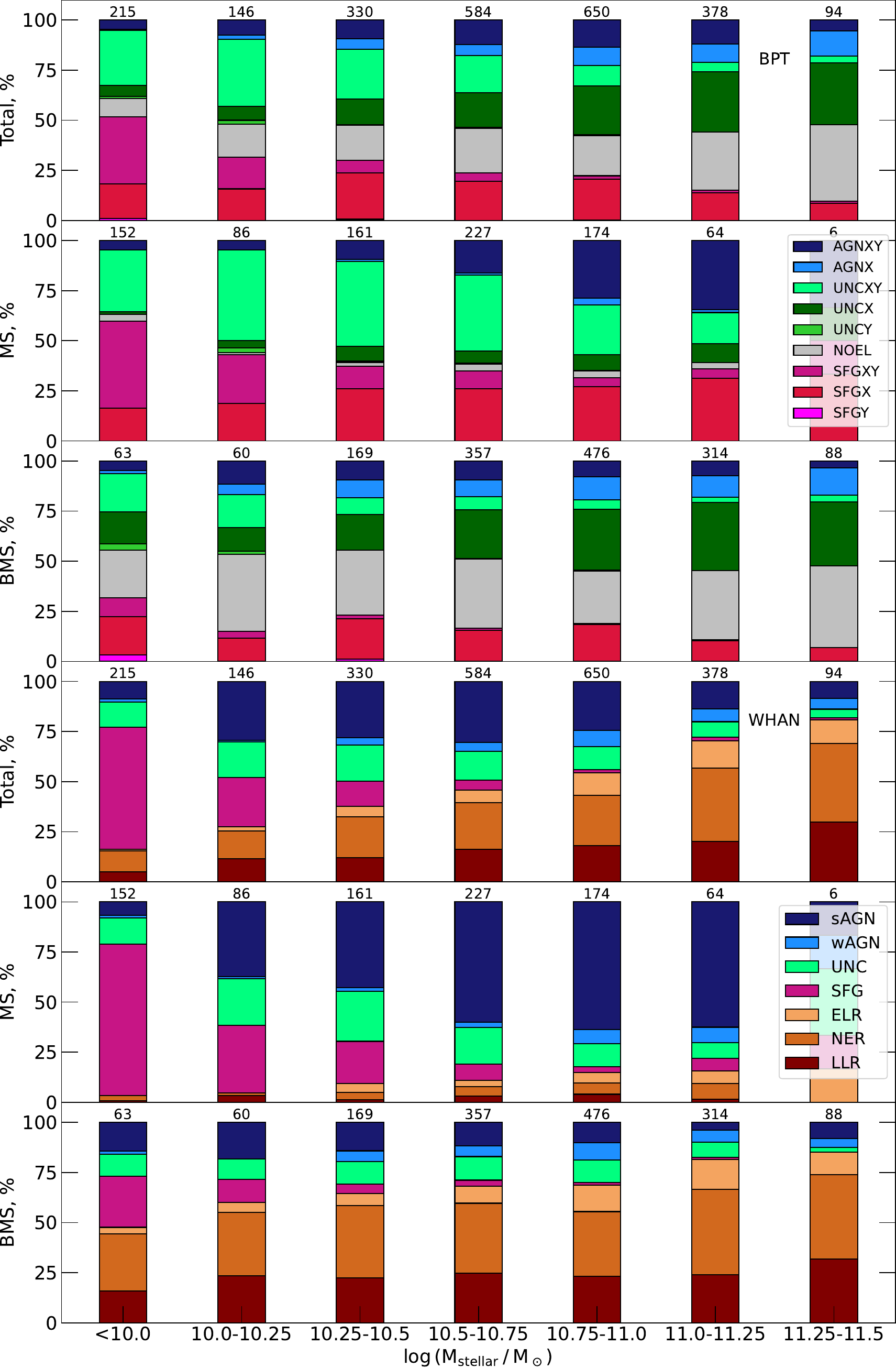}
\caption{BPT (top three panels) and the WHAN (bottom three panels) spectral class distribution as a funciton of stellar mass. In each group of panels, the top panel presents the distribution of the whole studied sample, the middle corresponds to the MS subsample, and the bottom displays the below-MS subsample. The information on colors can be found in legends. \label{fig:MASS_BINS}}
\end{figure*}


This section is devoted to the stellar mass distribution of galaxies concerning the BPT and WHAN spectral classes, presented in Fig.~\ref{fig:MASS_BINS}. The order for panels is similar to that presented in Fig.~\ref{fig:AGE_BINS}.


The stellar mass distribution considering the BPT classification of the full sample (the top panel from the top, Fig.~\ref{fig:MASS_BINS}) shows a systematical growth with a stellar mass of the fractions of UNCX, NOEL, and also a slight increase for AGNX, while SFGXY and UNCXY tend to decline. SFGX and AGNXY do not possess any distinct correlations.

The WHAN classification of the whole sample (the fourth panel from the top, Fig.~\ref{fig:MASS_BINS}) shows that more massive galaxies tend to be RGs.
In the total sample, sAGNs reach their maximum of $\sim 30\%$ for $\log {\rm (M_{stellar}/M_\odot)=10.5 - 10.75}$, but then slightly drop to $\sim10\%$ in the last bin. Both BPT and WHAN SFGs drop almost to $0$ 
at $\log {\rm (M_{stellar}/M_\odot)> 10.75}$.
Finally, wAGNs and UNC galaxies do not have any clear correlations with stellar mass. The percentage of RGs grows up for higher stellar masses, which is an expected result (e.g. \citealt{Stasinska2015}). 

\subsection{Outflows}
\label{res:out}

\begin{figure*}
\centering
\includegraphics[width=0.9\textwidth]{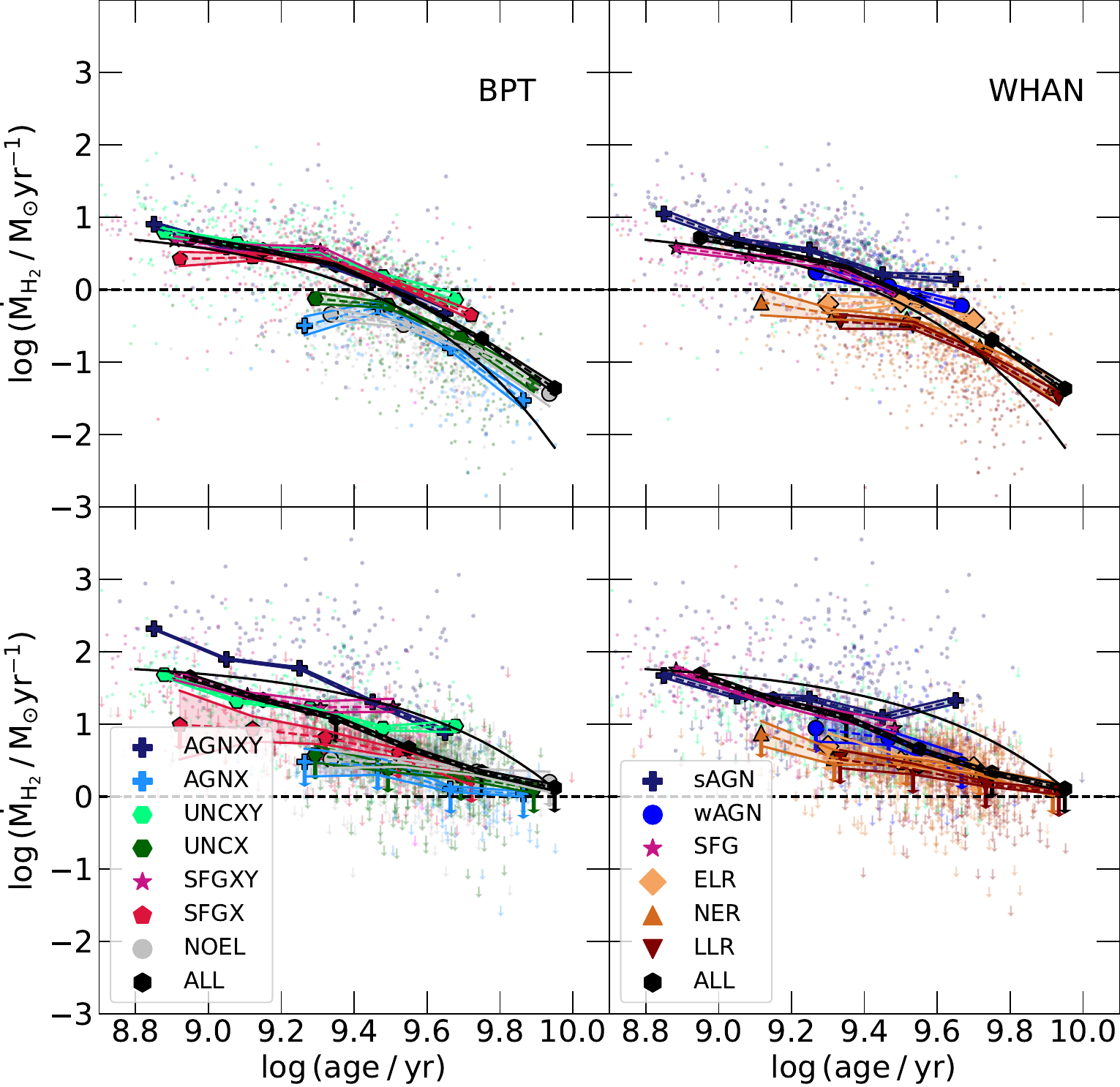}
\caption{Molecular hydrogen outflow rate \citep[eq. 5,][]{Fluetsch2019} as a function of light-weighted stellar age. The top panel displays outflow estimations only by stellar mass and star-formation rates, the bottom additionally considers $L_{\rm AGN} = 3500 L_{\textrm{\OOf}}$ \citep{Heckman2004}. Arrows at the bottom panel correspond to upper limits for cases of $2 \sigma$ non-detection of \OOf. Black solid lines are the fitted exponential decline functions to all galaxies: $\dot{M} [{\rm M_\odot \: yr^{-1}}] = 76.7 \pm 7.29 \: \exp(-t/[2.18 \pm 0.19] \; {\rm Gyr})$ (bottom panels), and $ \dot{M} [{\rm M_\odot \: yr^{-1}}] = 8.09 \pm 0.63 \: \exp(-t/[1.25 \pm 0.04] \; {\rm Gyr})$ (top panels).
\label{fig:OUT}}
\end{figure*}

\begin{figure*}
\centering
\includegraphics[width=0.95\textwidth]{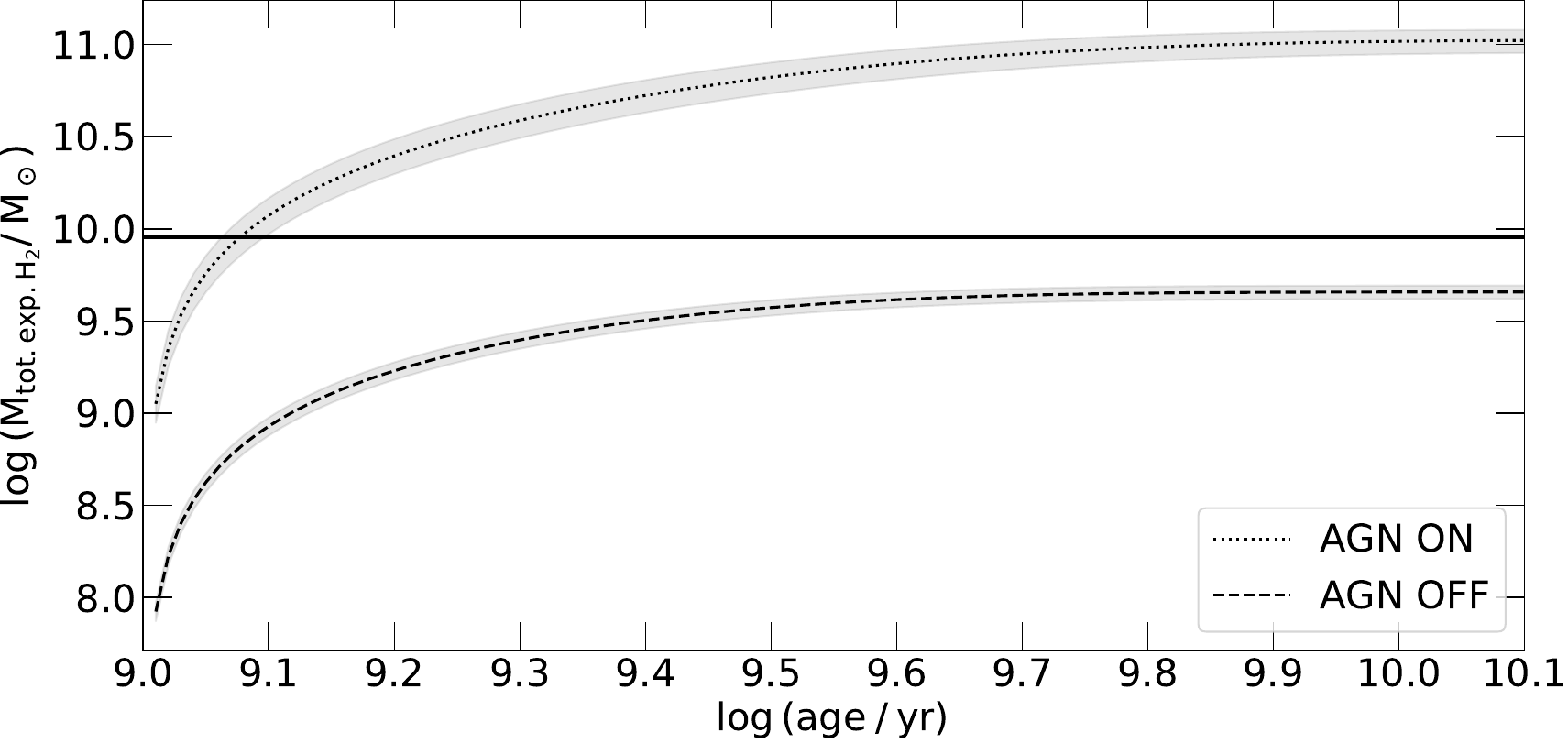}
\caption{The total expelled molecular hydrogen mass as a function of age, assuming the start of the outflow at $1 \: {\rm Gyr}$. The black dotted line represents the integration of the outflow function with considered $L_{\rm AGN}$ (`AGN on', eq.~\ref{func:agnon}), whereas the dashed line is based on the function without $L_{\rm AGN}$ (`AGN off', eq.~\ref{func:agnoff}) (See Section \ref{res:out}). The shaded regions around lines define $1\sigma$ uncertainties. The black horizontal line indicates the approximate expelled gas mass at $9 \: {\rm Gyr}$, based on CO-line observations of galaxies in the range between $1 \: {\rm Gyr}$ and $10 \: {\rm Gyr}$ \citep{Michalowski2024}. \label{fig:OUT_MASS}}
\end{figure*}

\begin{figure*}
\centering
\includegraphics[width=0.95\textwidth]{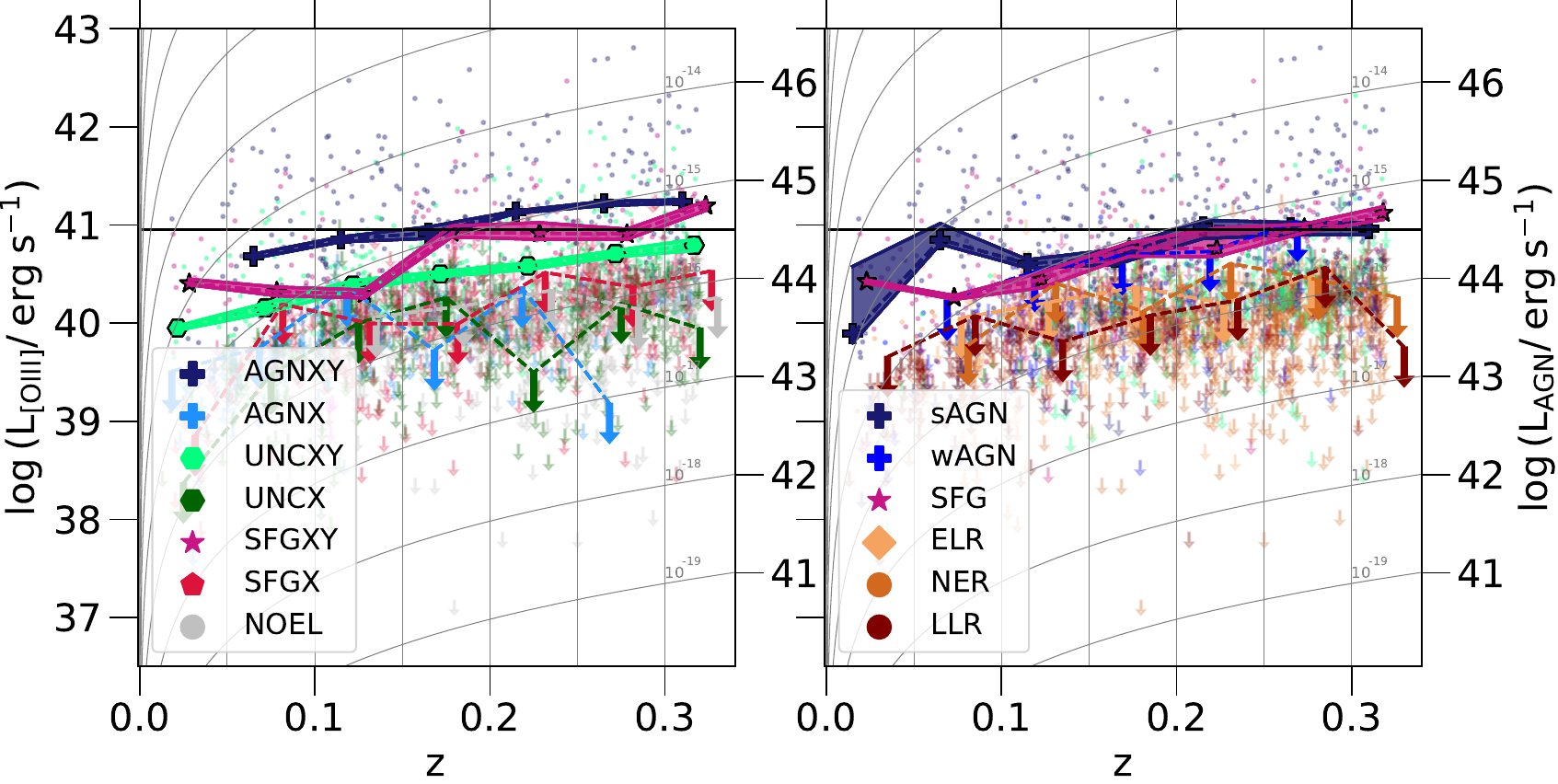}
\caption{The luminosity of the \OO line (left axis) and the estimated AGN luminosity (right axis), applying $L_{\rm AGN} = 3500 L_{\textrm{\OOf}}$ \citep{Heckman2004}, as a function of redshift, color-coded by the BPT (left panel) and WHAN (right panel) classes. Similarly to Fig.~\ref{fig:OUT}, the median points and median upper limits are shown, as the markers of corresponded color. The black solid line displays the typical AGN luminosity \citep{Roos2015}. The horizontal lines of the presented gray grid \label{fig:OIII} are the lines of an equal flux at different redshift, and the flux in ${\rm erg \: s^{-1} \: cm^{-2}}$ is written above each line. The vertical lines represent the redshift grid with the step of $0.05$.}
\end{figure*}

As in our previous paper in this series on a smaller sample \citep{Michalowski2024}, we analyzed the estimated outflow rates (eq. 5 in \citealt{Fluetsch2019}) as a function of age (Fig.~\ref{fig:OUT}). The outflow rates estimated using only stellar masses and star formation rates are presented at the top panel of Fig.~\ref{fig:OUT} (with the AGN luminosity set to zero, i.e.~`AGN off'), while the bottom panel displays the cases with additionally considered $L_{\rm AGN} = 3500 L_{\rm \text{\OOf}}$ (`AGN on') \citep{Heckman2004}. 

All panels of Fig.~\ref{fig:OUT} indicate that outflow rates decline exponentially with the age of galaxies. The fitted functions to `AGN on' (eq.~\ref{func:agnon}) and `AGN off' (eq.~\ref{func:agnoff}) estimations are the following:

\begin{equation}
    \label{func:agnon}
    \dot{M} [{\rm M_\odot \: yr^{-1}}] = 76.7 \pm 7.29 \: \exp(-t/[2.18 \pm 0.19] \; {\rm Gyr})
\end{equation}

\begin{equation}
    \label{func:agnoff}
    \dot{M} [{\rm M_\odot \: yr^{-1}}] = 8.09 \pm 0.63 \: \exp(-t/[1.25 \pm 0.04] \; {\rm Gyr})
\end{equation} 


Similarly to Section~\ref{res:temp}, we used a Monte-Carlo simulation to estimate the median values and their uncertainties with respect to the age bins and spectral classes. 
In order to obtain the correct medians, we also included upper limits in each age bin. 
If the percentage of upper limits in the subsample of galaxies exceeded 50$\%$, then the median also represents an upper limit (downward arrows in Fig.~\ref{fig:OUT}). Our findings suggest that AGN hosts and SFGs are likely to host powerful outflows ($10$ and $100\,{\rm M_\odot \: yr^{-1}}$ for `AGN-off' and `AGN-on' cases, respectively), especially for young galaxies, in contrast to old retired galaxies.

The consideration of AGN contribution brings out higher values of outflows, in several cases even reaching $\sim10^3 \: {\rm M_\odot yr^{-1}}$. Such high outflows would require only $10^7 - 10^8$ years to completely remove the ISM from a galaxy. This fact inspired us to conduct an average outflow analysis and estimate the timescale that is needed for a given outflow to fully remove the ISM from the galaxies.

Although we do not possess observational data for the currently studied sample to obtain the mass of molecular or atomic hydrogen, in the sample of \citet{Michalowski2024} the molecular gas mass declines from $\sim 10^{10} \: {\rm M_\odot}$ at $1 \: {\rm Gyr}$ to $\sim 10^9 \: {\rm M_\odot}$ at $9 \: {\rm  Gyr}$. Thus we decided to calculate the total expelled molecular gas mass by outflows by integrating the outflow rate functions and assuming the start of the outflow at $1 \: {\rm Gyr}$ (Fig.~\ref{fig:OUT_MASS}). According to Fig.~\ref{fig:OUT_MASS}, the role of $L_{\rm AGN}$ in the outflow estimation is decisive. The `AGN off' mode (dashed line) clearly is not capable of removing the observed gas mass at a realistic timescale, whereas the `AGN on' mode can remove up to $10^{11} \: {\rm M_\odot}$ of gas within the given time range, which is an order of magnitude higher than needed.

Lastly, Fig.~\ref{fig:OIII} presents the luminosity of \OO line and the AGN luminosity as a function of redshift. Analogously to Fig.~\ref{fig:OUT}, we calculated the medians and median upper limits to analyze the trends. Expectedly, the most luminous galaxies in all bins are BPT and WHAN AGN hosts, SFGs and UNC are slightly lower, and the lowest points are galaxies lacking emission lines (AGNX, UNCX, SFGX, and NOEL), and WHAN RGs. Both panels suggest, that the majority of points lying above the typical AGN luminosity line \citep{Roos2015}, are BPT AGNXY (135 galaxies, $58\%$) and WHAN sAGNs and wAGNs (164 galaxies, $70\%$).

\subsection{Environmental Effects}
\label{res:env}

\begin{figure*}
\centering
\includegraphics[width=\textwidth]{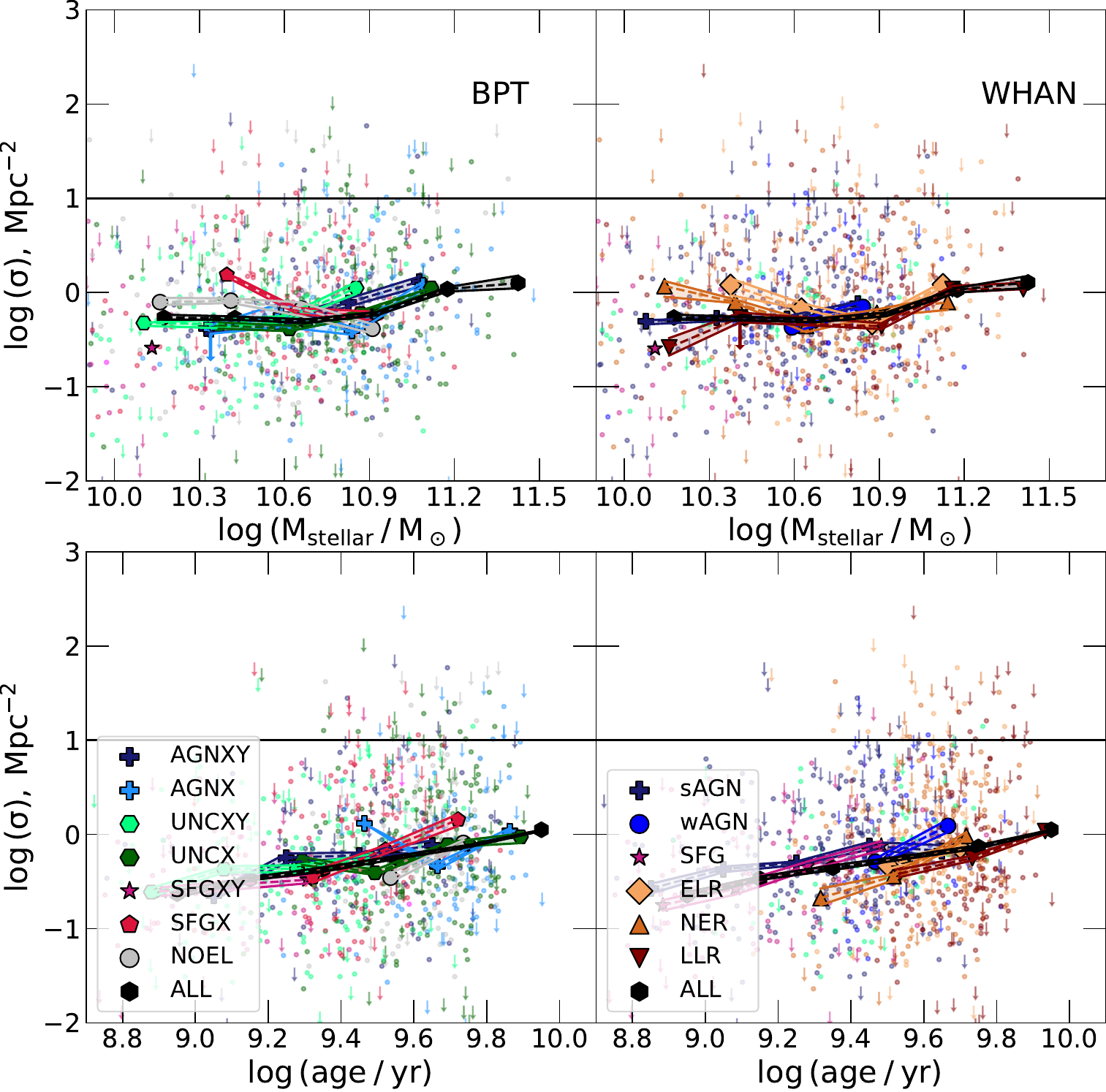}
\caption{The surface density, based
on the distance to the 5th nearest neighbor
in a velocity cylinder of $\pm1000 \: {\rm km \: s^{-1}}$ as a function of stellar mass (top panels) and age (bottom panels). Big markers with correspondent color represent the estimated median values (see Section~\ref{res:temp}) in each bin for BPT (left panels) and WHAN (right panels) spectral classes. The black solid line indicates the typical value of surface density for groups of galaxies.
\label{fig:SURFDEN}}
\end{figure*}

We explored the significance of the environmental effects on the ISM removal process. The GAMA catalog contains the measurement of the distance to the fifth neighbor 
\citep{Brough2013}, however, only 964 galaxies ($\sim 40\%$) in our sample have information about their environment. 
This does not bias our analysis, because the selection of these galaxies is only due to the redshift upper limit of $z<0.18$ imposed on this analysis in the GAMA catalog.

Fig.~\ref{fig:SURFDEN} displays the surface density based on the distance to the 5th nearest neighbor 
in a velocity cylinder of $\pm 1000 \: {\rm km \: s^{-1}}$ \footnote{\url{https://www.gama-survey.org/dr3/data/cat/EnvironmentMeasures/v05/}} as a function of stellar mass (top) and age (bottom). Upper limits either represent the galaxies that have an uncertainty in the surface density measurement higher than $50\%$, or cases for which no neighbors were found 
within the distance to the nearest angular survey edge, hence with the surface density value based on this distance.

As in previous sections, we divided the selected subsample into six age and mass bins and calculated median values (see Section \ref{res:temp}) in order to study possible trends. Upper limits were also included in the median estimation, and similarly to Section \ref{res:out}, if the percentage of upper limits exceeds $50\%$ for a spectral class in a bin, the calculated median is considered an upper limit, and marked with a downward arrow of a correspondent color. 

The galaxy density of $10~{\rm Mpc^{-2}}$ is a typical value of the surface density for a galaxy group \citep{Michalowski2024}, and is displayed as a solid horizontal line at all panels of Fig.~\ref{fig:SURFDEN}. The absolute majority of points are lying below the line, hence indicating that only $36$ galaxies out of $964$ are bounded in group structures. Although derived medians do not reveal a preference of these galaxies to reside in high-density environments for any of the spectral classes, the slight increase in galaxy number surface density for older galaxies can be observed for all spectral classes. 
However, this increase does not bring the older galaxies to high values of galaxy number surface density, indicating that the environment does not influence their ISM. 

\subsection{Machine-readable Table}

We created a machine-readable table (MRT), attached to this publication, with the data for all 305\,529 galaxies from the GAMA survey. The 5 rows with headers and descriptions from the MRT are presented in Table~\ref{tabMRT}. Briefly, the Table contains unique identifiers for each galaxy, such as its GAMA ID, right ascension and declination for J2000, and spectroscopic redshift. After that, the GAMA estimations of the S\'ersic index, its uncertainty, and the distance from the MS are listed. In the end, the available emission-line data for each galaxy was used to classify it, using our extended BPT and WHAN diagnostics (see Section~\ref{sec:method}), and also to calculate the outflow rate in the `AGN on' and `AGN off' modes (see Section~\ref{res:out}).  

\section{Discussion} 
\label{sec:disc}
ISM removal mechanisms may be divided into four groups by what they are powered by: AGN (Section~\ref{disc:AGN}), old stellar population (Section~\ref{disc:rgs}), recent star-formation processes and other phenomena i.e.~environmental effects, morphological quenching, etc. (Section~\ref{disc:other}). 

\subsection{The Role of the AGN Feedback in the ISM Removal Process}
\label{disc:AGN}
AGNs are believed to be one of the main mechanisms of ISM removal,
because of their significant feedback, e.g.~shocks, outflows, heating the gas to very high temperatures \citep[see][and reference therein for a review of the AGN feedback]{Fabian_2012}. By AGN feedback hereafter we consider heating, ionization, and outflows powered by AGNs.

Our key findings related to the role of the AGN feedback or the ISM removal process are the following: 

i) $23\%$ of the studied sample was classified as WHAN sAGN, while the BPT classification suggested $11\%$ AGNs (Fig.~\ref{fig:BMSMS}). Nevertheless, the number of secure optical AGN hosts (classified as such by both BPT and WHAN, adopting the conditions of \citealt{Riffel_2023}) in our sample is $7\%$.

ii) Median cold dust temperatures of WHAN sAGNs are $1-4 \: {\rm K}$ higher for the age bins up to $\log({\rm age}/{\rm yr})=9.4$ than those of WHAN SFGs in the same age bin. The BPT AGNXY and SFGXY show the same tendency, although the temperature difference is lower. Both diagnostics also suggest that warm dust temperatures of AGNs stay within the uncertainties with those of SFGs (Fig.~\ref{fig:TEMP}).

iii) The MS subsample consists of 16\% BPT AGNXY and 46\% of WHAN sAGNs, while the below-MS subsample possesses only 8\% of AGNXY and 10\% of sAGNs, thus indicating that AGNs tend to appear in MS galaxies (Fig.~\ref{fig:BMSMS}), and on average they have SFRs at the same level as for SFGs (Fig.~\ref{fig:SFRSM})

iv) The fraction of WHAN sAGNs in age bins decreases, starting from the age of $10^{9.0}$ yr (Fig.~\ref{fig:AGE_BINS}). The slight decline is also observable for the BPT AGNXY category, but it is vital to note that $\sim20\%$ of the AGNXY galaxies are classified as WHAN RGs (see Appendix~\ref{ax:corr}), with this fraction increasing for higher age bins.

These results support a scenario with ionization and heating by the detectable AGN playing a substantial role only for the galaxies with $\log {\rm (age/yr) < 9.4}$. We observe the decline of the fraction of BPT AGNXY and WHAN sAGNs in age bins above $10^{9.2}$ yr. AGN hosts tend to have higher cold dust temperatures than non-AGNs up to the age of $\log {\rm (age/yr)} = 9.4$, both for BPT and WHAN. In addition to that, median cold dust temperatures of AGN hosts for both classifications continuously decline (Fig.~\ref{fig:TEMP}). However, the rate of the ISM removal is not decreasing for higher ages, because the dust decline is consistent with an exponential function (Fig.~\ref{fig:MDMS} and \citealt{M19,Lesniewska2023,Michalowski2024}). The fraction of AGNs decreases with age, so, even taking into account the duty cycle effect (see below), this suggests some other mechanism than AGN feedback being dominant for galaxies with ages $10^{9.4} - 10^{10.0}$ yr in our sample. The contribution of the old stellar population, particularly HOLMES, supporting the ISM removal in the old passive galaxies, might be the solution, which is discussed in Section~\ref{disc:rgs}.

Spectrally classified AGN hosts also show similar values of SFRs to the SFGs with similar stellar masses, which might be interpreted as AGNs not affecting the star-formation process. 
However, the MAGPHYS estimations of SFR do not take into consideration the fact that a part of the emission might originate from AGN, so the SFRs of AGN hosts might be overestimated. Lower SFRs of BPT AGNXY than those of SFGXY and UNCXY galaxies might be explained by the fact that $\sim20\%$ of BPT AGNXY are WHAN RGs (see Appendix~\ref{ax:corr}), mostly those in the LINER region (top panels of Fig.~\ref{fig:BPT_WHAN}). 

As discussed in \citet{Michalowski2024}, we need to consider the fact that supermassive black holes are active only for a small fraction of time, so we do not expect all recent AGNs currently showing AGN spectral signatures. Due to this duty cycle phenomena \citep{Novak_2011, Hickox_2014, Padovani_2017, Hardcastle2018}, the sample of galaxies dominated by AGNs 
would show the current AGN activity in $1-10\%$ of galaxies, depending on the Eddington ratio. 


The duty cycle suggests that most of the galaxies in our sample
might have been affected by the AGN feedback at some point. The effect of
this past feedback would not show up in the spectral analysis presented in this
work. This however would imply a roughly constant fraction of AGN with
age. Figure~\ref{fig:AGE_BINS} shows however the opposite. The AGN fraction decreases with
age, while the RG fraction increases. This evolutionary sequence might be
explained with the following, non-exclusive scenarios: 1) ionization by HOLMES is surpassing that of AGN, becoming more and more dominant for higher ages; 2) AGNs cease, i.e.~they stay less and less time in the active phase or become less luminous, which might be potentially caused by the lack of cold gas in older galaxies \citep{Spilker2018, Michalowski2024}. This would lead to the lack of matter around the SMBH to accrete, and hence, would not allow the SMBH to enter the AGN phase. Either way, this implies a diminishing role of the AGN feedback in the ISM removal process for galaxies with ages of $10^{9.4} - 10^{10.0}$\,yr.

Any of the above scenarios is consistent with our general conclusion that ionization and temperature feedback from AGNs play a crucial role in the ISM removal only in galaxies with ages up to $\log{\rm (age/yr) < 9.4}$ in the studied sample. However, the ISM removal for older galaxies cannot be powered mainly by the AGN feedback.

We note that this conclusion may seem to contradict the results of \citet{Nadolny2024}, where it was shown that the AGN radio mode accretion may effectively prevent gas from cooling at ages $\log{\rm (age/yr) > 9.4}$. However, that result concerns preventing gas cooling, not removing the existing cold gas reservoir.

An important caveat of our analysis is connected with the AGN luminosity detection threshold. The measured AGN fraction depends on the detection threshold of the data we use. We show this using the \OO luminosities and upper limits as a function of redshift on Fig.~\ref{fig:OIII}. We note that most of our upper limits are below the bolometric luminosity of $10^{44.5}\,\mbox{erg}\,\mbox{s}^{-1}$ (solid line), which is usually regarded as a typical AGN luminosity \citep{Roos2015}. Hence, our data is sensitive to all but weak AGNs. Hence, our data cannot be used to reject the hypothesis that weak AGNs are solely responsible for ISM removal.

We also note that the fact that the AGN fraction decreases with age is not a selection effect. First, our estimate of age is based on the broad-band SED modeling independent of spectral line analysis. 

Second, the detectability of AGNs should actually improve (or at least be the same) for older galaxies. This is because the stellar and ISM components of galaxies at the same redshift and with the same stellar mass are fainter for older galaxies than for younger galaxies at all relevant emission lines and blue continuum, and similar at the red continuum, so the dilution effect is weaker, and consequently a lower-luminosity AGN can be detected.
The BPT and WHAN diagrams are only biased against dust-obscured AGNs, but they are unlikely to be common in old galaxies. Moreover, our mid-infrared analysis, sensitive to dusty AGNs, does not reveal many AGNs (out of galaxies neither classified as AGN hosts via the BPT nor via WHAN there are only 26 and 32 WISE AGNs applying the \citealt{Stern2012} and \citealt{Mateos2012} thresholds, respectively).

We also estimated the rates of outflows powered by AGNs or starbursts 
(see Section~\ref{res:out}), because
they
may be capable of expelling large amounts of gas from galaxies. Therefore, outflows might play an essential role in the ISM removal process, and, consequently, in the quenching of galaxies.

According to Fig.~\ref{fig:OUT}, setting $L_{\rm AGN}$ to zero (`AGN off' mode), estimated outflow rates are too low to explain the ISM removal exclusively. However, applying $L_{\rm AGN} = 3500 L_{\textrm{\OOf}}$ (`AGN on' mode) results in substantially higher outflow rates in optical AGN hosts (and some SFGs). These might be interpreted as outflows playing a significant role in removing the ISM in galaxies showing current AGN activity, which exhibit higher median outflow rate values than any other spectral classes. On the other hand, the estimated outflow rates for RGs and BPT classes lacking the detection of emission lines (AGNX, UNCX, NOEL) are 1-2 orders of magnitude lower than those of AGNs in the same age bin, and in many cases are below the limit for sufficient average outflow of ${\rm 1 \: M_{\rm{\odot}}/yr}$.

The anomalously high outflow rates, and hence, short expected timescales of the full ISM removal might be explained by the already discussed duty cycle of the AGN.
AGNs stay in a passive mode for $\sim90\%$ of time, during which they are not energetic enough to power up substantial outflows. 
AGNs in the active phase might expel a substantial amount of the ISM, but for a short time.


\subsection{The Nature of Retired Galaxies}
\label{disc:rgs}

RGs are galaxies with quenched star formation and gas ionized by HOLMES, i.e.~low-mass stars in post-AGB/planetary nebulae (PNe) phase. These stars are also suspected to explain the LI(N)ER emission in ETGs, thus placing the galaxy above the AGN delimitation line at the BPT diagram \citep[see][and reference therein for a review of HOLMES and RGs]{Stasinska_2008, Cid_Fernandes_2011, Herpich_2018}.


The WHAN diagnostic reveals that nearly $50\%$ of our sample is classified as different classes of RGs: with weak emission lines (ELR, $8\%$), with absorption or undetected lines (LLR, $16\%$), and with noisy emission lines (NER, $25\%$)  (Fig.~\ref{fig:BMSMS}). RGs tend to lie below the MS (Fig.~\ref{fig:SFRSM}), hence populate the below-MS subsample (Fig.~\ref{fig:BMSMS}), which is expected, due to the fact, that RGs possess either little star-formation or even complete absence of it, but also tend to be massive (Fig.~\ref{fig:MASS_BINS}), which places them below the MS. The $10\%$ of MS subsample being classified as RGs (Fig.~\ref{fig:BMSMS}) is due to retired galaxies located slightly above the separation line between MS and below-MS subsamples at the SFR-M\* plot (Fig.~\ref{fig:SFRSM}), hence, this is just selection criteria effect. Figure~\ref{fig:AGE_BINS} presents that the percentage of RGs grows significantly for higher ages, which correlates with their definition \citep{Cid_Fernandes_2010, Herpich_2016}. 
The temperature analysis (Fig.~\ref{fig:TEMP}) demonstrated that RGs are the coldest galaxies among the spectral classes, however, their cold dust temperature tends to stay at the same level with age, unlike sAGNs.


\citet{Herpich_2018} suggested that both LLR and ELR galaxies have similar ionizing photon budgets, produced by HOLMES, and the only difference in their nature is the presence of a warm gas reservoir in ELR galaxies. Those findings also align with the analysis of \citet{Lorenzoni2024}.

The median cold dust temperatures of ELR are $\sim1 \: {\rm K}$ higher than NER and LLR (see Fig. \ref{fig:TEMP}), which might be interpreted as the presence of a warm gas reservoir in ELR. However, we do not have any clues on the origin of this warm gas reservoir: \citet{Herpich_2018} concluded that ELRs have recently experienced a star-forming episode ignited by an inflow of gas that was either left around the galaxy after the merger episode \citep{Rutherford2024} or previously expelled by strong stellar winds.


The presented results might be interpreted as HOLMES playing an exceptional role in the ISM removal process, being the dominant ionization source for $\sim50\%$ of our sample, and for an even higher fraction of galaxies in age bins $9.4 < \log {\rm (age/yr)} <  10.0$ and with the stellar masses $\log{\rm(M_{stellar}/M_{\odot})>10.75}$. The constant temperature medians might be interpreted as HOLMES being an effective source of heat, supporting the thermal equilibrium.

This provides another piece of evidence for planetary nebulae/HOLMES being viable mechanisms of ISM removal. The spectral analysis is not able to reveal the supernovae Ia (SNe Ia) contribution, thus we are not able to test the conclusion of \citet{Michalowski2024}. However, the simulations of ETGs indicate the insignificance of SNe Ia in the ISM removal \citep{Nadolny2024}.

\subsection{Other Mechanisms of ISM Removal}
\label{disc:other}

In \citet{Michalowski2024} it is shown that there are three main possible mechanisms of the ISM removal: ionization by planetary nebulae, and ionization/outflows by supernovae Type Ia and AGNs. Several possible scenarios (e.g.~astration, environmental effects) were excluded, based on the observational evidence. However, in this article, we perform an analysis of the bigger sample, so we also test the role of previously excluded scenarios.

\begin{enumerate}
    

    \item {\em Astration and feedback from young massive stars.} Despite the fact that $\sim7\%$ of the studied sample was classified as SFGs by both BPT and WHAN, their SFRs on average are too low to expect any substantial effect of either consuming the ISM for forming stars or ionizing it by young massive stars \citep{Michalowski2024}. In addition, if these mechanisms played an essential role in the ISM removal, we should reveal a slower dust decline for higher light-weighted stellar ages \citep{M19, Michalowski2024}, which is not the case, according to Fig~\ref{fig:MDMS}. However, the ionization and heating by young massive stars might be an additional mechanism to the AGN feedback that removes the ISM in the younger galaxies of the studied sample (Fig.~\ref{fig:AGE_BINS}).

    
    \item {\em Environmental effects.} Environmental effects might play an important role in the ISM removal process due to gravitational interaction between galaxies, heating processes in the cluster environment, ram pressure stripping etc.~\citep{Jachym2012, Mok2016}.

    The analysis of the surface number density of galaxies presented in Section \ref{res:env} indicated the weak trend that all spectral classes tend to be in slightly denser environments with higher ages. However, the estimated median values are an order of magnitude lower, than the group threshold. 
    Therefore, environmental effects might occur in several galaxies in the studied sample, however, they are not able to explain the ISM removal process in all of tem. 

\end{enumerate}

\section{Conclusions} \label{sec:conc}

We analyze optical spectra and physical parameters of 2409 dusty early-type galaxies. We present extended BPT and WHAN spectral diagnostics, which allowed us to classify all galaxies from the studied sample by their dominant ionization source, independently of the signal-to-noise limit of the required emission lines. 

That classification and further analysis of physical parameters (e.g.~dust temperatures, star-formation rates etc.) revealed that ionization, heating and outflows, powered by detectable AGNs can play a major role in the ISM removal only for galaxies with ages $\log{\rm (age/yr) < 9.4}$, however for older galaxies the AGN activity ceases and is probably replaced by the HOLMES and PN contribution. However, we cannot rule out the influence of weak AGNs at any age. The ISM removal mechanisms caused by the recent star-formation process (i.e.~ionization and heating by young massive stars, astration) may affect only the youngest galaxies in the studied sample, but, as we showed in our previous articles, they are not sufficient to explain gas and dust removal \citep{M19, Michalowski2024}. Finally, we show that less than $4\%$ of the galaxies with available environmental measurements 
are likely bounded in group structures and, thus might host current environmental effects, which rules out the environmental effects as a dominant mechanism of ISM removal.

Additionally, we analyzed several potential biases that might alter AGN classification and our extension of BPT and WHAN diagnostics, which are discussed in detail in the Appendix. We present our extended BPT and WHAN classifications for 300\,000 galaxies in the GAMA fields.




\begin{acknowledgements}

O.R.~acknowledges the support of 
the National Science Centre, Poland through the grant 2022/01/4/ST9/00037.
M.J.M., J.N., and M.S.~acknowledge the support of 
the National Science Centre, Poland through the SONATA BIS grant 2018/30/E/ST9/00208.
M.J.M.~acknowledges the support of
the Polish National Agency for Academic Exchange (NAWA) Bekker grant BPN/BEK/2022/1/00110. J.N.~acknowledges the support of the NAWA Bekker grant  BPN/BEK/2023/1/00271.
This research was funded in whole or in part by the National Science Centre, Poland, grant number 2023/49/B/ST9/00066 
and 2021/41/N/ST9/02662. 
For the purpose of Open Access, the author has applied a CC-BY public copyright license to any Author Accepted Manuscript (AAM) version arising from this submission.
J.H.~and A.L.~were supported by a VILLUM FONDEN Investigator grant (project number 16599). This work was supported by a research grant (VIL54489) from VILLUM FONDEN. This article is based upon work from COST Action CA21126 - Carbon molecular nanostructures in space (NanoSpace), supported by COST (European Cooperation in Science and Technology). This work has been supported by the Japan Society for the Promotion of Science (JSPS) Grants-in-Aid for Scientific Research (19H05076, 21H01128, and 24H00247).
This work has also been supported in part by the Sumitomo Foundation Fiscal 2018 Grant for Basic Science Research Projects (180923), and the Collaboration Funding of the Institute of Statistical Mathematics ``New Development of the Studies on Galaxy Evolution with a Method of Data Science''. C.G. is supported by a VILLUM FONDEN Young Investigator grant (project number 25501).

\end{acknowledgements}

\bibliography{main}{}

\begin{thebibliography}{}
\expandafter\ifx\csname natexlab\endcsname\relax\def\natexlab#1{#1}\fi
\providecommand{\url}[1]{\href{#1}{#1}}
\providecommand{\dodoi}[1]{doi:~\href{http://doi.org/#1}{\nolinkurl{#1}}}
\providecommand{\doeprint}[1]{\href{http://ascl.net/#1}{\nolinkurl{http://ascl.net/#1}}}
\providecommand{\doarXiv}[1]{\href{https://arxiv.org/abs/#1}{\nolinkurl{https://arxiv.org/abs/#1}}}

\bibitem[{{Agostino} \& {Salim}(2019)}]{Agostino2019}
{Agostino}, C.~J., \& {Salim}, S. 2019, \apj, 876, 12, \dodoi{10.3847/1538-4357/ab1094}

\bibitem[{{Agostino} {et~al.}(2023){Agostino}, {Salim}, {Ellison}, {Bickley}, \& {Faber}}]{Agostino2023_SDSS}
{Agostino}, C.~J., {Salim}, S., {Ellison}, S.~L., {Bickley}, R.~W., \& {Faber}, S.~M. 2023, \apj, 943, 174, \dodoi{10.3847/1538-4357/acac99}

\bibitem[{{Alb{\'a}n} \& {Wylezalek}(2023)}]{Alban2023}
{Alb{\'a}n}, M., \& {Wylezalek}, D. 2023, \aap, 674, A85, \dodoi{10.1051/0004-6361/202245437}

\bibitem[{{Baldry} {et~al.}(2018){Baldry}, {Liske}, {Brown}, {Robotham}, {Driver}, {Dunne}, {Alpaslan}, {Brough}, {Cluver}, {Eardley}, {Farrow}, {Heymans}, {Hildebrandt}, {Hopkins}, {Kelvin}, {Loveday}, {Moffett}, {Norberg}, {Owers}, {Taylor}, {Wright}, {Bamford}, {Bland -Hawthorn}, {Bourne}, {Bremer}, {Colless}, {Conselice}, {Croom}, {Davies}, {Foster}, {Grootes}, {Holwerda}, {Jones}, {Kafle}, {Kuijken}, {Lara-Lopez}, {L{\'o}pez-S{\'a}nchez}, {Meyer}, {Phillipps}, {Sutherland}, {van Kampen}, \& {Wilkins}}]{Baldry2018}
{Baldry}, I.~K., {Liske}, J., {Brown}, M.~J.~I., {et~al.} 2018, MNRAS, 474, 3875, \dodoi{10.1093/mnras/stx3042}

\bibitem[{{Baldwin} {et~al.}(1981){Baldwin}, {Phillips}, \& {Terlevich}}]{bpt}
{Baldwin}, J.~A., {Phillips}, M.~M., \& {Terlevich}, R. 1981, \pasp, 93, 5, \dodoi{10.1086/130766}

\bibitem[{{Brough} {et~al.}(2013){Brough}, {Croom}, {Sharp}, {Hopkins}, {Taylor}, {Baldry}, {Gunawardhana}, {Liske}, {Norberg}, {Robotham}, {Bauer}, {Bland-Hawthorn}, {Colless}, {Foster}, {Kelvin}, {Lara-Lopez}, {L{\'o}pez-S{\'a}nchez}, {Loveday}, {Owers}, {Pimbblet}, \& {Prescott}}]{Brough2013}
{Brough}, S., {Croom}, S., {Sharp}, R., {et~al.} 2013, \mnras, 435, 2903, \dodoi{10.1093/mnras/stt1489}

\bibitem[{{Cid Fernandes} {et~al.}(2011){Cid Fernandes}, {Stasi{\'n}ska}, {Mateus}, \& {Vale Asari}}]{Cid_Fernandes_2011}
{Cid Fernandes}, R., {Stasi{\'n}ska}, G., {Mateus}, A., \& {Vale Asari}, N. 2011, \mnras, 413, 1687, \dodoi{10.1111/j.1365-2966.2011.18244.x}

\bibitem[{{Cid Fernandes} {et~al.}(2010){Cid Fernandes}, {Stasi{\'n}ska}, {Schlickmann}, {Mateus}, {Vale Asari}, {Schoenell}, \& {Sodr{\'e}}}]{Cid_Fernandes_2010}
{Cid Fernandes}, R., {Stasi{\'n}ska}, G., {Schlickmann}, M.~S., {et~al.} 2010, \mnras, 403, 1036, \dodoi{10.1111/j.1365-2966.2009.16185.x}

\bibitem[{{da Cunha} {et~al.}(2008){da Cunha}, {Charlot}, \& {Elbaz}}]{Cuhna2008}
{da Cunha}, E., {Charlot}, S., \& {Elbaz}, D. 2008, \mnras, 388, 1595, \dodoi{10.1111/j.1365-2966.2008.13535.x}

\bibitem[{{de Vaucouleurs} {et~al.}(1978){de Vaucouleurs}, {de Vaucouleurs}, \& {Corwin}}]{Vaucouleurs1978}
{de Vaucouleurs}, G., {de Vaucouleurs}, A., \& {Corwin}, H.~G. 1978, \aj, 83, 1331, \dodoi{10.1086/112322}

\bibitem[{{Dekel} \& {Silk}(1986)}]{Dekel1986}
{Dekel}, A., \& {Silk}, J. 1986, \apj, 303, 39, \dodoi{10.1086/164050}

\bibitem[{{Driver} {et~al.}(2011){Driver}, {Hill}, {Kelvin}, {Robotham}, {Liske}, {Norberg}, {Baldry}, {Bamford}, {Hopkins}, {Loveday}, {Peacock}, {Andrae}, {Bland-Hawthorn}, {Brough}, {Brown}, {Cameron}, {Ching}, {Colless}, {Conselice}, {Croom}, {Cross}, {de Propris}, {Dye}, {Drinkwater}, {Ellis}, {Graham}, {Grootes}, {Gunawardhana}, {Jones}, {van Kampen}, {Maraston}, {Nichol}, {Parkinson}, {Phillipps}, {Pimbblet}, {Popescu}, {Prescott}, {Roseboom}, {Sadler}, {Sansom}, {Sharp}, {Smith}, {Taylor}, {Thomas}, {Tuffs}, {Wijesinghe}, {Dunne}, {Frenk}, {Jarvis}, {Madore}, {Meyer}, {Seibert}, {Staveley-Smith}, {Sutherland}, \& {Warren}}]{Driver2011}
{Driver}, S.~P., {Hill}, D.~T., {Kelvin}, L.~S., {et~al.} 2011, \mnras, 413, 971, \dodoi{10.1111/j.1365-2966.2010.18188.x}

\bibitem[{{Driver} {et~al.}(2016){Driver}, {Wright}, {Andrews}, {Davies}, {Kafle}, {Lange}, {Moffett}, {Mannering}, {Robotham}, {Vinsen}, {Alpaslan}, {Andrae}, {Baldry}, {Bauer}, {Bamford}, {Bland-Hawthorn}, {Bourne}, {Brough}, {Brown}, {Cluver}, {Croom}, {Colless}, {Conselice}, {da Cunha}, {De Propris}, {Drinkwater}, {Dunne}, {Eales}, {Edge}, {Frenk}, {Graham}, {Grootes}, {Holwerda}, {Hopkins}, {Ibar}, {van Kampen}, {Kelvin}, {Jarrett}, {Jones}, {Lara-Lopez}, {Liske}, {Lopez-Sanchez}, {Loveday}, {Maddox}, {Madore}, {Mahajan}, {Meyer}, {Norberg}, {Penny}, {Phillipps}, {Popescu}, {Tuffs}, {Peacock}, {Pimbblet}, {Prescott}, {Rowlands}, {Sansom}, {Seibert}, {Smith}, {Sutherland}, {Taylor}, {Valiante}, {Vazquez-Mata}, {Wang}, {Wilkins}, \& {Williams}}]{Driver2016}
{Driver}, S.~P., {Wright}, A.~H., {Andrews}, S.~K., {et~al.} 2016, MNRAS, 455, 3911, \dodoi{10.1093/mnras/stv2505}

\bibitem[{{Driver} {et~al.}(2022){Driver}, {Bellstedt}, {Robotham}, {Baldry}, {Davies}, {Liske}, {Obreschkow}, {Taylor}, {Wright}, {Alpaslan}, {Bamford}, {Bauer}, {Bland-Hawthorn}, {Bilicki}, {Bravo}, {Brough}, {Casura}, {Cluver}, {Colless}, {Conselice}, {Croom}, {de Jong}, {D'Eugenio}, {De Propris}, {Dogruel}, {Drinkwater}, {Dvornik}, {Farrow}, {Frenk}, {Giblin}, {Graham}, {Grootes}, {Gunawardhana}, {Hashemizadeh}, {H{\"a}u{\ss}ler}, {Heymans}, {Hildebrandt}, {Holwerda}, {Hopkins}, {Jarrett}, {Heath Jones}, {Kelvin}, {Koushan}, {Kuijken}, {Lara-L{\'o}pez}, {Lange}, {L{\'o}pez-S{\'a}nchez}, {Loveday}, {Mahajan}, {Meyer}, {Moffett}, {Napolitano}, {Norberg}, {Owers}, {Radovich}, {Raouf}, {Peacock}, {Phillipps}, {Pimbblet}, {Popescu}, {Said}, {Sansom}, {Seibert}, {Sutherland}, {Thorne}, {Tuffs}, {Turner}, {van der Wel}, {van Kampen}, \& {Wilkins}}]{Driver2022}
{Driver}, S.~P., {Bellstedt}, S., {Robotham}, A. S.~G., {et~al.} 2022, \mnras, 513, 439, \dodoi{10.1093/mnras/stac472}

\bibitem[{{Fabian}(2012)}]{Fabian_2012}
{Fabian}, A.~C. 2012, \araa, 50, 455, \dodoi{10.1146/annurev-astro-081811-125521}

\bibitem[{{Federrath} \& {Klessen}(2012)}]{Federrath2012}
{Federrath}, C., \& {Klessen}, R.~S. 2012, \apj, 761, 156, \dodoi{10.1088/0004-637X/761/2/156}

\bibitem[{{Fluetsch} {et~al.}(2019){Fluetsch}, {Maiolino}, {Carniani}, {Marconi}, {Cicone}, {Bourne}, {Costa}, {Fabian}, {Ishibashi}, \& {Venturi}}]{Fluetsch2019}
{Fluetsch}, A., {Maiolino}, R., {Carniani}, S., {et~al.} 2019, \mnras, 483, 4586, \dodoi{10.1093/mnras/sty3449}

\bibitem[{{Gensior} {et~al.}(2020){Gensior}, {Kruijssen}, \& {Keller}}]{Gensior2020}
{Gensior}, J., {Kruijssen}, J.~M.~D., \& {Keller}, B.~W. 2020, \mnras, 495, 199, \dodoi{10.1093/mnras/staa1184}

\bibitem[{{Gordon} {et~al.}(2017){Gordon}, {Owers}, {Pimbblet}, {Croom}, {Alpaslan}, {Baldry}, {Brough}, {Brown}, {Cluver}, {Conselice}, {Davies}, {Holwerda}, {Hopkins}, {Gunawardhana}, {Loveday}, {Taylor}, \& {Wang}}]{Gordon2017}
{Gordon}, Y.~A., {Owers}, M.~S., {Pimbblet}, K.~A., {et~al.} 2017, \mnras, 465, 2671, \dodoi{10.1093/mnras/stw2925}

\bibitem[{{Griffin} {et~al.}(2010){Griffin}, {Abergel}, {Abreu}, {Ade}, {Andr{\'e}}, {Augueres}, {Babbedge}, {Bae}, {Baillie}, {Baluteau}, {Barlow}, {Bendo}, {Benielli}, {Bock}, {Bonhomme}, {Brisbin}, {Brockley-Blatt}, {Caldwell}, {Cara}, {Castro-Rodriguez}, {Cerulli}, {Chanial}, {Chen}, {Clark}, {Clements}, {Clerc}, {Coker}, {Communal}, {Conversi}, {Cox}, {Crumb}, {Cunningham}, {Daly}, {Davis}, {de Antoni}, {Delderfield}, {Devin}, {di Giorgio}, {Didschuns}, {Dohlen}, {Donati}, {Dowell}, {Dowell}, {Duband}, {Dumaye}, {Emery}, {Ferlet}, {Ferrand}, {Fontignie}, {Fox}, {Franceschini}, {Frerking}, {Fulton}, {Garcia}, {Gastaud}, {Gear}, {Glenn}, {Goizel}, {Griffin}, {Grundy}, {Guest}, {Guillemet}, {Hargrave}, {Harwit}, {Hastings}, {Hatziminaoglou}, {Herman}, {Hinde}, {Hristov}, {Huang}, {Imhof}, {Isaak}, {Israelsson}, {Ivison}, {Jennings}, {Kiernan}, {King}, {Lange}, {Latter}, {Laurent}, {Laurent}, {Leeks}, {Lellouch}, {Levenson}, {Li}, {Li}, {Lilienthal}, {Lim}, {Liu}, {Lu}, {Madden}, {Mainetti}, {Marliani},
  {McKay}, {Mercier}, {Molinari}, {Morris}, {Moseley}, {Mulder}, {Mur}, {Naylor}, {Nguyen}, {O'Halloran}, {Oliver}, {Olofsson}, {Olofsson}, {Orfei}, {Page}, {Pain}, {Panuzzo}, {Papageorgiou}, {Parks}, {Parr-Burman}, {Pearce}, {Pearson}, {P{\'e}rez-Fournon}, {Pinsard}, {Pisano}, {Podosek}, {Pohlen}, {Polehampton}, {Pouliquen}, {Rigopoulou}, {Rizzo}, {Roseboom}, {Roussel}, {Rowan-Robinson}, {Rownd}, {Saraceno}, {Sauvage}, {Savage}, {Savini}, {Sawyer}, {Scharmberg}, {Schmitt}, {Schneider}, {Schulz}, {Schwartz}, {Shafer}, {Shupe}, {Sibthorpe}, {Sidher}, {Smith}, {Smith}, {Smith}, {Spencer}, {Stobie}, {Sudiwala}, {Sukhatme}, {Surace}, {Stevens}, {Swinyard}, {Trichas}, {Tourette}, {Triou}, {Tseng}, {Tucker}, {Turner}, {Vaccari}, {Valtchanov}, {Vigroux}, {Virique}, {Voellmer}, {Walker}, {Ward}, {Waskett}, {Weilert}, {Wesson}, {White}, {Whitehouse}, {Wilson}, {Winter}, {Woodcraft}, {Wright}, {Xu}, {Zavagno}, {Zemcov}, {Zhang}, \& {Zonca}}]{spire}
{Griffin}, M.~J., {Abergel}, A., {Abreu}, A., {et~al.} 2010, \aap, 518, L3, \dodoi{10.1051/0004-6361/201014519}

\bibitem[{{Hardcastle}(2018)}]{Hardcastle2018}
{Hardcastle}, M.~J. 2018, \mnras, 475, 2768, \dodoi{10.1093/mnras/stx3358}

\bibitem[{{Heckman}(1980)}]{Heckman1980}
{Heckman}, T.~M. 1980, \aap, 87, 152

\bibitem[{{Heckman} {et~al.}(2004){Heckman}, {Kauffmann}, {Brinchmann}, {Charlot}, {Tremonti}, \& {White}}]{Heckman2004}
{Heckman}, T.~M., {Kauffmann}, G., {Brinchmann}, J., {et~al.} 2004, \apj, 613, 109, \dodoi{10.1086/422872}

\bibitem[{{Herpich} {et~al.}(2016){Herpich}, {Mateus}, {Stasi{\'n}ska}, {Cid Fernandes}, \& {Vale Asari}}]{Herpich_2016}
{Herpich}, F., {Mateus}, A., {Stasi{\'n}ska}, G., {Cid Fernandes}, R., \& {Vale Asari}, N. 2016, \mnras, 462, 1826, \dodoi{10.1093/mnras/stw1742}

\bibitem[{{Herpich} {et~al.}(2018){Herpich}, {Stasi{\'n}ska}, {Mateus}, {Vale Asari}, \& {Cid Fernandes}}]{Herpich_2018}
{Herpich}, F., {Stasi{\'n}ska}, G., {Mateus}, A., {Vale Asari}, N., \& {Cid Fernandes}, R. 2018, \mnras, 481, 1774, \dodoi{10.1093/mnras/sty2391}

\bibitem[{{Hickox} {et~al.}(2014){Hickox}, {Mullaney}, {Alexander}, {Chen}, {Civano}, {Goulding}, \& {Hainline}}]{Hickox_2014}
{Hickox}, R.~C., {Mullaney}, J.~R., {Alexander}, D.~M., {et~al.} 2014, \apj, 782, 9, \dodoi{10.1088/0004-637X/782/1/9}

\bibitem[{{Ho}(1999)}]{Ho1999}
{Ho}, L.~C. 1999, Advances in Space Research, 23, 813, \dodoi{10.1016/S0273-1177(99)00211-2}

\bibitem[{{Iglesias-P{\'a}ramo} {et~al.}(2016){Iglesias-P{\'a}ramo}, {V{\'\i}lchez}, {Rosales-Ortega}, {S{\'a}nchez}, {Duarte Puertas}, {Petropoulou}, {Gil de Paz}, {Galbany}, {Moll{\'a}}, {Catal{\'a}n-Torrecilla}, {Castillo Morales}, {Mast}, {Husemann}, {Garc{\'\i}a-Benito}, {Mendoza}, {Kehrig}, {P{\'e}rez-Montero}, {Papaderos}, {Gomes}, {Walcher}, {Gonz{\'a}lez Delgado}, {Marino}, {L{\'o}pez-S{\'a}nchez}, {Ziegler}, {Flores}, \& {Alves}}]{IglesiasParamo2016}
{Iglesias-P{\'a}ramo}, J., {V{\'\i}lchez}, J.~M., {Rosales-Ortega}, F.~F., {et~al.} 2016, \apj, 826, 71, \dodoi{10.3847/0004-637X/826/1/71}

\bibitem[{{J{\'a}chym} \& {Palou{\v{s}}}(2012)}]{Jachym2012}
{J{\'a}chym}, P., \& {Palou{\v{s}}}, J. 2012, in EAS Publications Series, Vol.~56, EAS Publications Series, ed. M.~A. {de Avillez}, 113--117, \dodoi{10.1051/eas/1256015}

\bibitem[{{Kauffmann} {et~al.}(2003){Kauffmann}, {Heckman}, {Tremonti}, {Brinchmann}, {Charlot}, {White}, {Ridgway}, {Brinkmann}, {Fukugita}, {Hall}, {Ivezi{\'c}}, {Richards}, \& {Schneider}}]{Kauffmann2003}
{Kauffmann}, G., {Heckman}, T.~M., {Tremonti}, C., {et~al.} 2003, \mnras, 346, 1055, \dodoi{10.1111/j.1365-2966.2003.07154.x}

\bibitem[{{Kelvin} {et~al.}(2012){Kelvin}, {Driver}, {Robotham}, {Hill}, {Alpaslan}, {Baldry}, {Bamford}, {Bland-Hawthorn}, {Brough}, {Graham}, {H{\"a}ussler}, {Hopkins}, {Liske}, {Loveday}, {Norberg}, {Phillipps}, {Popescu}, {Prescott}, {Taylor}, \& {Tuffs}}]{Kelvin2012}
{Kelvin}, L.~S., {Driver}, S.~P., {Robotham}, A. S.~G., {et~al.} 2012, \mnras, 421, 1007, \dodoi{10.1111/j.1365-2966.2012.20355.x}

\bibitem[{{Kewley} {et~al.}(2001){Kewley}, {Dopita}, {Sutherland}, {Heisler}, \& {Trevena}}]{Kewley2001}
{Kewley}, L.~J., {Dopita}, M.~A., {Sutherland}, R.~S., {Heisler}, C.~A., \& {Trevena}, J. 2001, \apj, 556, 121, \dodoi{10.1086/321545}

\bibitem[{{Kewley} {et~al.}(2006){Kewley}, {Groves}, {Kauffmann}, \& {Heckman}}]{Kewley2006}
{Kewley}, L.~J., {Groves}, B., {Kauffmann}, G., \& {Heckman}, T. 2006, \mnras, 372, 961, \dodoi{10.1111/j.1365-2966.2006.10859.x}

\bibitem[{{Le{\'s}niewska} {et~al.}(2023){Le{\'s}niewska}, {Micha{\l}owski}, {Gall}, {Hjorth}, {Nadolny}, {Ryzhov}, \& {Solar}}]{Lesniewska2023}
{Le{\'s}niewska}, A., {Micha{\l}owski}, M.~J., {Gall}, C., {et~al.} 2023, \apj, 953, 27, \dodoi{10.3847/1538-4357/acdcfc}

\bibitem[{{Lorenzoni} {et~al.}(2024){Lorenzoni}, {Rembold}, \& {de Carvalho}}]{Lorenzoni2024}
{Lorenzoni}, V., {Rembold}, S.~B., \& {de Carvalho}, R.~R. 2024, \mnras, 527, 3542, \dodoi{10.1093/mnras/stad3435}

\bibitem[{{Maragkoudakis} {et~al.}(2014){Maragkoudakis}, {Zezas}, {Ashby}, \& {Willner}}]{Maragkoudakis2014}
{Maragkoudakis}, A., {Zezas}, A., {Ashby}, M.~L.~N., \& {Willner}, S.~P. 2014, \mnras, 441, 2296, \dodoi{10.1093/mnras/stu634}

\bibitem[{{Martig} {et~al.}(2009){Martig}, {Bournaud}, {Teyssier}, \& {Dekel}}]{Martig2009}
{Martig}, M., {Bournaud}, F., {Teyssier}, R., \& {Dekel}, A. 2009, \apj, 707, 250, \dodoi{10.1088/0004-637X/707/1/250}

\bibitem[{{Mateos} {et~al.}(2012){Mateos}, {Alonso-Herrero}, {Carrera}, {Blain}, {Watson}, {Barcons}, {Braito}, {Severgnini}, {Donley}, \& {Stern}}]{Mateos2012}
{Mateos}, S., {Alonso-Herrero}, A., {Carrera}, F.~J., {et~al.} 2012, \mnras, 426, 3271, \dodoi{10.1111/j.1365-2966.2012.21843.x}

\bibitem[{{Micha{\l}owski} {et~al.}(2019){Micha{\l}owski}, {Hjorth}, {Gall}, {Frayer}, {Tsai}, {Hirashita}, {Rowlands}, {Takeuchi}, {Le{\'s}niewska}, {Behrendt}, {Bourne}, {Hughes}, {Spring}, {Zavala}, \& {Bartczak}}]{M19}
{Micha{\l}owski}, M.~J., {Hjorth}, J., {Gall}, C., {et~al.} 2019, \aap, 632, A43, \dodoi{10.1051/0004-6361/201936055}

\bibitem[{{Micha{\l}owski} {et~al.}(2024){Micha{\l}owski}, {Gall}, {Hjorth}, {Frayer}, {Tsai}, {Rowlands}, {Takeuchi}, {Le{\'s}niewska}, {Behrendt}, {Bourne}, {Hughes}, {Koprowski}, {Nadolny}, {Ryzhov}, {Solar}, {Spring}, {Zavala}, \& {Bartczak}}]{Michalowski2024}
{Micha{\l}owski}, M.~J., {Gall}, C., {Hjorth}, J., {et~al.} 2024, \apj, 964, 129, \dodoi{10.3847/1538-4357/ad1b52}

\bibitem[{{Mok} \& {Wilson}(2016)}]{Mok2016}
{Mok}, A., \& {Wilson}, C. 2016, in From Interstellar Clouds to Star-Forming Galaxies: Universal Processes?, ed. P.~{Jablonka}, P.~{Andr{\'e}}, \& F.~{van der Tak}, Vol. 315, E56, \dodoi{10.1017/S1743921316008188}

\bibitem[{{Muratov} {et~al.}(2015){Muratov}, {Kere{\v{s}}}, {Faucher-Gigu{\`e}re}, {Hopkins}, {Quataert}, \& {Murray}}]{Muratov2015}
{Muratov}, A.~L., {Kere{\v{s}}}, D., {Faucher-Gigu{\`e}re}, C.-A., {et~al.} 2015, \mnras, 454, 2691, \dodoi{10.1093/mnras/stv2126}

\bibitem[{{Nadolny} {et~al.}(2024){Nadolny}, {Micha{\l}owski}, {Parente}, {Hjorth}, {Gall}, {Le{\'s}niewska}, {Solar}, {Nowaczyk}, \& {Ryzhov}}]{Nadolny2024}
{Nadolny}, J., {Micha{\l}owski}, M.~J., {Parente}, M., {et~al.} 2024, \aap, 689, A210, \dodoi{10.1051/0004-6361/202449839}

\bibitem[{{Novak} {et~al.}(2011){Novak}, {Ostriker}, \& {Ciotti}}]{Novak_2011}
{Novak}, G.~S., {Ostriker}, J.~P., \& {Ciotti}, L. 2011, \apj, 737, 26, \dodoi{10.1088/0004-637X/737/1/26}

\bibitem[{{Padoan} \& {Nordlund}(2002)}]{Padoan2002}
{Padoan}, P., \& {Nordlund}, {\r{A}}. 2002, \apj, 576, 870, \dodoi{10.1086/341790}

\bibitem[{{Padovani} {et~al.}(2017){Padovani}, {Alexander}, {Assef}, {De Marco}, {Giommi}, {Hickox}, {Richards}, {Smol{\v{c}}i{\'c}}, {Hatziminaoglou}, {Mainieri}, \& {Salvato}}]{Padovani_2017}
{Padovani}, P., {Alexander}, D.~M., {Assef}, R.~J., {et~al.} 2017, \aapr, 25, 2, \dodoi{10.1007/s00159-017-0102-9}

\bibitem[{{Peng} {et~al.}(2015){Peng}, {Maiolino}, \& {Cochrane}}]{Peng2015}
{Peng}, Y., {Maiolino}, R., \& {Cochrane}, R. 2015, \nat, 521, 192, \dodoi{10.1038/nature14439}

\bibitem[{{Peterson}(2006)}]{Peterson2006}
{Peterson}, B.~M. 2006, in Physics of Active Galactic Nuclei at all Scales, ed. D.~{Alloin}, Vol. 693, 77, \dodoi{10.1007/3-540-34621-X_3}

\bibitem[{{Piotrowska} {et~al.}(2022){Piotrowska}, {Bluck}, {Maiolino}, \& {Peng}}]{Piotrowska2022}
{Piotrowska}, J.~M., {Bluck}, A. F.~L., {Maiolino}, R., \& {Peng}, Y. 2022, \mnras, 512, 1052, \dodoi{10.1093/mnras/stab3673}

\bibitem[{{Riffel} {et~al.}(2023){Riffel}, {Mallmann}, {Rembold}, {Ilha}, {Riffel}, {Storchi-Bergmann}, {Ruschel-Dutra}, {Vazdekis}, {Mart{\'\i}n-Navarro}, {Schimoia}, {Ramos Almeida}, {da Costa}, {Vila-Verde}, \& {Gatto}}]{Riffel_2023}
{Riffel}, R., {Mallmann}, N.~D., {Rembold}, S.~B., {et~al.} 2023, \mnras, 524, 5640, \dodoi{10.1093/mnras/stad2234}

\bibitem[{{Roos} {et~al.}(2015){Roos}, {Juneau}, {Bournaud}, \& {Gabor}}]{Roos2015}
{Roos}, O., {Juneau}, S., {Bournaud}, F., \& {Gabor}, J.~M. 2015, \apj, 800, 19, \dodoi{10.1088/0004-637X/800/1/19}

\bibitem[{{Rutherford} {et~al.}(2024){Rutherford}, {van de Sande}, {Croom}, {Valenzuela}, {Remus}, {D'Eugenio}, {Vaughan}, {Zovaro}, {Casura}, {Barsanti}, {Bland-Hawthorn}, {Brough}, {Bryant}, {Goodwin}, {Lorente}, {Oh}, \& {Ristea}}]{Rutherford2024}
{Rutherford}, T.~H., {van de Sande}, J., {Croom}, S.~M., {et~al.} 2024, \mnras, 529, 810, \dodoi{10.1093/mnras/stae398}

\bibitem[{{Saintonge} \& {Catinella}(2022)}]{Saintonge2022}
{Saintonge}, A., \& {Catinella}, B. 2022, \araa, 60, 319, \dodoi{10.1146/annurev-astro-021022-043545}

\bibitem[{{S{\'a}nchez} {et~al.}(2015){S{\'a}nchez}, {P{\'e}rez}, {Rosales-Ortega}, {Miralles-Caballero}, {L{\'o}pez-S{\'a}nchez}, {Iglesias-P{\'a}ramo}, {Marino}, {S{\'a}nchez-Menguiano}, {Garc{\'\i}a-Benito}, {Mast}, {Mendoza}, {Papaderos}, {Ellis}, {Galbany}, {Kehrig}, {Monreal-Ibero}, {Gonz{\'a}lez Delgado}, {Moll{\'a}}, {Ziegler}, {de Lorenzo-C{\'a}ceres}, {Mendez-Abreu}, {Bland-Hawthorn}, {Bekerait{\.{e}}}, {Roth}, {Pasquali}, {D{\'\i}az}, {Bomans}, {van de Ven}, \& {Wisotzki}}]{Sanchez2015}
{S{\'a}nchez}, S.~F., {P{\'e}rez}, E., {Rosales-Ortega}, F.~F., {et~al.} 2015, \aap, 574, A47, \dodoi{10.1051/0004-6361/201424873}

\bibitem[{{Sazonova} {et~al.}(2021){Sazonova}, {Alatalo}, {Rowlands}, {Deustua}, {French}, {Heckman}, {Lanz}, {Lisenfeld}, {Luo}, {Medling}, {Nyland}, {Otter}, {Petric}, {Snyder}, \& {Urry}}]{Sazonova2021}
{Sazonova}, E., {Alatalo}, K., {Rowlands}, K., {et~al.} 2021, \apj, 919, 134, \dodoi{10.3847/1538-4357/ac0f7f}

\bibitem[{{Schawinski} {et~al.}(2014){Schawinski}, {Urry}, {Simmons}, {Fortson}, {Kaviraj}, {Keel}, {Lintott}, {Masters}, {Nichol}, {Sarzi}, {Skibba}, {Treister}, {Willett}, {Wong}, \& {Yi}}]{Schawinski2014}
{Schawinski}, K., {Urry}, C.~M., {Simmons}, B.~D., {et~al.} 2014, \mnras, 440, 889, \dodoi{10.1093/mnras/stu327}

\bibitem[{{S{\'e}rsic}(1963)}]{Sersic1963}
{S{\'e}rsic}, J.~L. 1963, Boletin de la Asociacion Argentina de Astronomia La Plata Argentina, 6, 41

\bibitem[{{Smith} {et~al.}(2011){Smith}, {Dunne}, {Maddox}, {Eales}, {Bonfield}, {Jarvis}, {Sutherland}, {Fleuren}, {Rigby}, {Thompson}, {Baldry}, {Bamford}, {Buttiglione}, {Cava}, {Clements}, {Cooray}, {Croom}, {Dariush}, {de Zotti}, {Driver}, {Dunlop}, {Fritz}, {Hill}, {Hopkins}, {Hopwood}, {Ibar}, {Ivison}, {Jones}, {Kelvin}, {Leeuw}, {Liske}, {Loveday}, {Madore}, {Norberg}, {Panuzzo}, {Pascale}, {Pohlen}, {Popescu}, {Prescott}, {Robotham}, {Rodighiero}, {Scott}, {Seibert}, {Sharp}, {Temi}, {Tuffs}, {van der Werf}, \& {van Kampen}}]{Smith2011}
{Smith}, D.~J.~B., {Dunne}, L., {Maddox}, S.~J., {et~al.} 2011, \mnras, 416, 857, \dodoi{10.1111/j.1365-2966.2011.18827.x}

\bibitem[{{Speagle} {et~al.}(2014){Speagle}, {Steinhardt}, {Capak}, \& {Silverman}}]{Speagle_2014}
{Speagle}, J.~S., {Steinhardt}, C.~L., {Capak}, P.~L., \& {Silverman}, J.~D. 2014, \apjs, 214, 15, \dodoi{10.1088/0067-0049/214/2/15}

\bibitem[{{Spilker} {et~al.}(2018){Spilker}, {Bezanson}, {Bari{\v{s}}i{\'c}}, {Bell}, {Lagos}, {Maseda}, {Muzzin}, {Pacifici}, {Sobral}, {Straatman}, {van der Wel}, {van Dokkum}, {Weiner}, {Whitaker}, {Williams}, \& {Wu}}]{Spilker2018}
{Spilker}, J., {Bezanson}, R., {Bari{\v{s}}i{\'c}}, I., {et~al.} 2018, \apj, 860, 103, \dodoi{10.3847/1538-4357/aac438}

\bibitem[{{Stasi{\'n}ska} {et~al.}(2006){Stasi{\'n}ska}, {Cid Fernandes}, {Mateus}, {Sodr{\'e}}, \& {Asari}}]{Stasinska2006}
{Stasi{\'n}ska}, G., {Cid Fernandes}, R., {Mateus}, A., {Sodr{\'e}}, L., \& {Asari}, N.~V. 2006, \mnras, 371, 972, \dodoi{10.1111/j.1365-2966.2006.10732.x}

\bibitem[{{Stasi{\'n}ska} {et~al.}(2015){Stasi{\'n}ska}, {Costa-Duarte}, {Vale Asari}, {Cid Fernandes}, \& {Sodr{\'e}}}]{Stasinska2015}
{Stasi{\'n}ska}, G., {Costa-Duarte}, M.~V., {Vale Asari}, N., {Cid Fernandes}, R., \& {Sodr{\'e}}, L. 2015, \mnras, 449, 559, \dodoi{10.1093/mnras/stv078}

\bibitem[{{Stasi{\'n}ska} {et~al.}(2008){Stasi{\'n}ska}, {Vale Asari}, {Cid Fernandes}, {Gomes}, {Schlickmann}, {Mateus}, {Schoenell}, {Sodr{\'e}}, \& {Seagal Collaboration}}]{Stasinska_2008}
{Stasi{\'n}ska}, G., {Vale Asari}, N., {Cid Fernandes}, R., {et~al.} 2008, \mnras, 391, L29, \dodoi{10.1111/j.1745-3933.2008.00550.x}

\bibitem[{{Stern} {et~al.}(2012){Stern}, {Assef}, {Benford}, {Blain}, {Cutri}, {Dey}, {Eisenhardt}, {Griffith}, {Jarrett}, {Lake}, {Masci}, {Petty}, {Stanford}, {Tsai}, {Wright}, {Yan}, {Harrison}, \& {Madsen}}]{Stern2012}
{Stern}, D., {Assef}, R.~J., {Benford}, D.~J., {et~al.} 2012, \apj, 753, 30, \dodoi{10.1088/0004-637X/753/1/30}

\bibitem[{{Veilleux} {et~al.}(1995){Veilleux}, {Kim}, {Sanders}, {Mazzarella}, \& {Soifer}}]{Veilleux1995}
{Veilleux}, S., {Kim}, D.~C., {Sanders}, D.~B., {Mazzarella}, J.~M., \& {Soifer}, B.~T. 1995, \apjs, 98, 171, \dodoi{10.1086/192158}

\bibitem[{{Wylezalek} {et~al.}(2022){Wylezalek}, {Cicone}, {Belfiore}, {Bertemes}, {Cazzoli}, {Wagg}, {Wang (王无忌)}, {Aravena}, {Maiolino}, {Martin}, {Bothwell}, {Brownstein}, {Bundy}, \& {De Breuck}}]{Wylezalek2022}
{Wylezalek}, D., {Cicone}, C., {Belfiore}, F., {et~al.} 2022, \mnras, 510, 3119, \dodoi{10.1093/mnras/stab3356}

\end{thebibliography}
\bibliographystyle{aasjournal}

\appendix

\section{Correspondence between extended BPT and WHAN diagnostics}
\label{ax:corr}

\begin{figure*}
\centering
\includegraphics[width=0.9\textwidth]{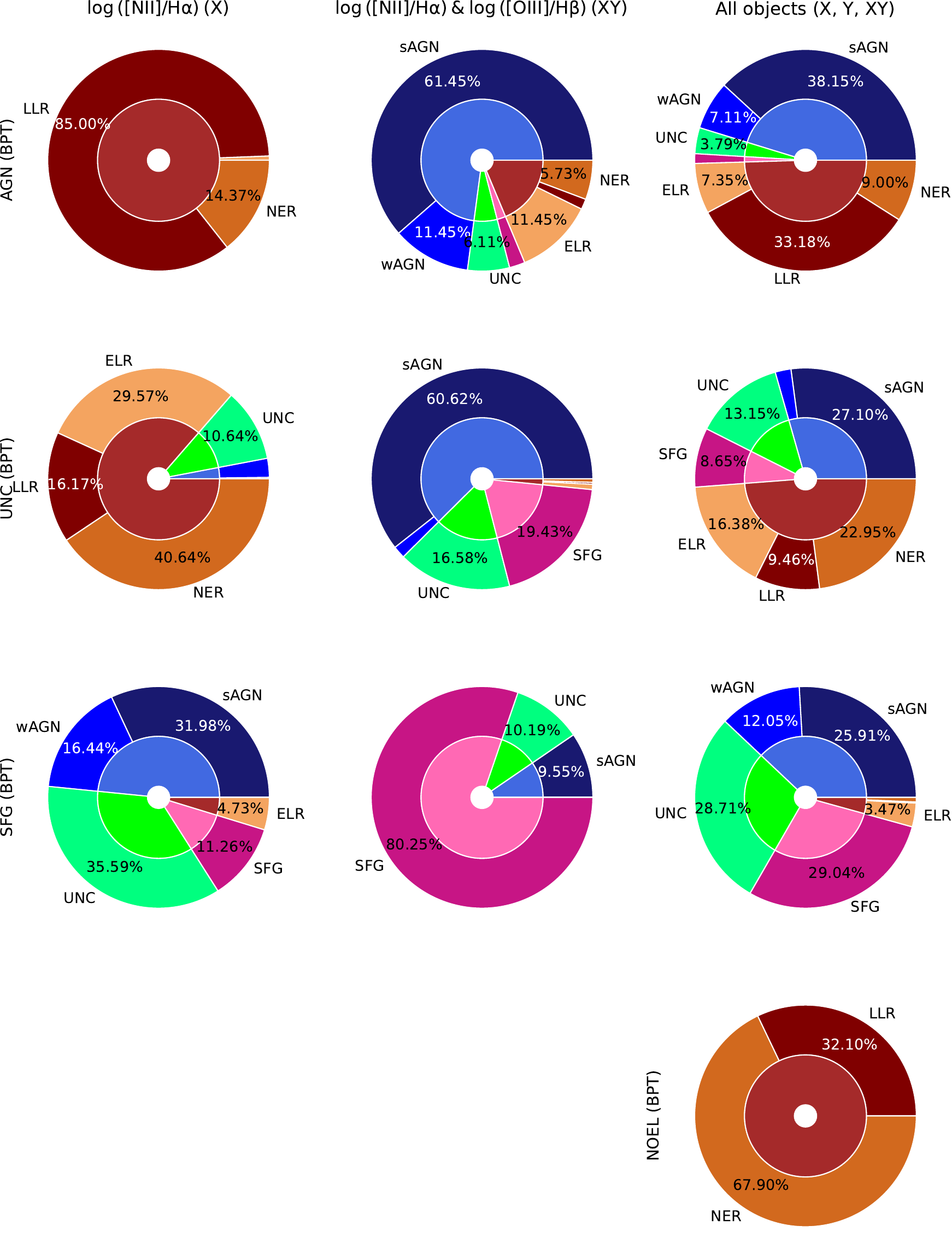}
\caption{The WHAN spectral classification of the subsamples of BPT AGN (top panels), UNC (second row of panels from the top), SFG (third row of panels from the top), and NOEL (bottom panels) galaxies. The left panels demonstrate the galaxies, classified only using the X-axis (AGNX, UNCX, SFX; see Section \ref{BPTALGO}), the middle panels - both axes (XY-objects); and the right panels present results for all objects of the correspondent BPT class (X, Y and XY). Similarly to Fig.~\ref{fig:BMSMS}, the outer circle displays detailed information on the spectral classes, whereas general classification into AGN, SFG, UNC, and RGs is represented in the inner circle.}
\label{fig:WHANvsBPT}
\end{figure*}

The main advantage of the extended BPT and WHAN diagnostics presented in this work is the absence of statistical bias, caused by the signal-to-noise cut of emission lines. Below, we discuss the special features of this extension and other explicit advantages of this method.

We provided a few additional classes for the BPT classification (NOEL, AGNX, SFGX, UNCX, AGNY, etc.) to solve the problem of non-detections of emission lines. We have only 16 galaxies with no detections of ${\rm H\alpha}$ or \NNf, but with detected \OO and/or ${\rm H\beta}$ lines (the Y-objects). The small fraction of Y-objects may be explained by the fact that \OO and ${\rm H\beta}$ require more energetic photons than \NN and ${\rm H\alpha}$, respectively. Thus, if a galaxy lacks a detection of ${\rm H\alpha}$ and \NNf, it most likely also lacks a detection of \OO and ${\rm H\beta}$.

Conversely, X-objects (i.e.~classified only via the \NNf$\mr{/H\alpha}$ ratio) constitute approximately $45\%$ of our sample (Fig.~\ref{fig:BMSMS}), playing a significant role in our conclusions.

In order to examine the correspondence between both diagnostics, we show the WHAN classification for each BPT class separately, excluding Y-objects due to their low number. The results of this analysis are displayed in Fig.~\ref{fig:WHANvsBPT}.

The top panels of Fig.~\ref{fig:WHANvsBPT} show the distributions of the WHAN classes for the BPT AGNX  (left), AGNXY (middle), and all AGN classes (right). The AGNX class is dominated by retired galaxies (RGs), with a negligible fraction of sAGN and wAGN. On the other hand, AGNXY are classified as RGs in only $18\%$ of cases, while sAGN and wAGN classes account for $73\%$. Therefore, among all BPT-selected AGNs there are $45\%$ WHAN AGNs, $50\%$  WHAN RGs, and less than $4\%$ SFGs and UNC. We also report that $29\%$ of WHAN sAGN and $24\%$ of WHAN wAGN galaxies are classified as AGN by the BPT diagram.

The second row of panels from the top has the same structure as the top one but shows unclear BPT cases. The overwhelming majority ($\sim85\%$) of UNCX is classified as RGs, leaving $15\%$ of galaxies classified as WHAN wAGN and UNC. The UNCXY galaxies show a different distribution: $62\%$ are WHAN sAGNs and wAGNs, $19\%$ are SFGs, and $17\%$ are WHAN UNC classes, leaving less than $3\%$ for RGs. The BPT UNC category is a mix of all WHAN classes with $49\%$ RGs, $29\%$ AGNs, $9\%$ SFG, and $13\%$ WHAN unclear cases.

The third row from the top presents BPT SFGs. It is clear that SFGXs are classified as SFGs by WHAN in $11\%$ of cases, tending to be rather classified as UNC ($36\%$), sAGN and wAGN ($48\%$), and ELR ($5\%$). However, $80\%$ of SFGXY galaxies are classified as WHAN SFGs, and the remaining $20\%$ are classified as UNC and sAGN. Overall, the distribution of the WHAN spectral classes for BPT SFGs is the following: WHAN classified $29\%$ as SFG, $29\%$ as UNC, $38\%$ as AGNs, and less than $5\%$ of RGs.

The bottom panel contains information about the NOEL, which consists of, as expected, only NERs and LLR galaxies. We did not perform the analysis of AGNY, UNCY, and SFGY due to the low number of galaxies in these categories, see Table~\ref{tabBPT}.

Figure~\ref{fig:WHANvsBPT} shows that AGNX and UNCX objects are mostly classified as RGs. This can be explained by the weak or undetected ${\rm H\alpha}$ line and detected \NNf, which places these galaxies below the SFG line of the WHAN diagram. The large number of sAGN and wAGN in the SFGX category can be explained by different values of the delimiter between SFG and AGN: $\log($\NNf$\mr{/H\alpha}) = -0.3$ for the WHAN diagram and $\log($\NNf$\mr{/H\alpha}) < 0.05$ for the BPT diagram.

The top middle panel of Fig.~\ref{fig:BPT_WHAN} shows the position of galaxies on the BPT diagram, color-coded by the classes from the WHAN diagram.
Systematically, the BPT LINERs region, defined by \citet{Cid_Fernandes_2011}, is filled with ELR galaxies.

Thus, we stress the fact that AGNX and UNCX objects in our classification should be considered in the overwhelming majority as RGs, lacking strong emission lines. 

\section{Aperture effects}
\label{ax:apeff}

Recent high-resolution IFU spectroscopy surveys of galaxies (e.g.~MaNGA, CALIFA, etc.) have led to further discussion of the misclassification of AGNs by single-aperture spectroscopic surveys (e.g.~GAMA, SDSS, etc.) due to dilution by star-forming gas \citep{Riffel_2023}, or to saturation by diffuse ionized gas (DIG) emission \citep{Alban2023}. Although the most recent comparison of X-ray observations and optical spectroscopic data from SDSS of galaxies with redshift $z \sim 0.1$ by \citet{Agostino2023_SDSS} has shown that such effect is negligible, we decided to inspect the dependency between spectral classes on both redshift and effective radius, that might hint on the bias caused by a fixed aperture size.

\begin{figure*}[p!]
\centering
\includegraphics[width=0.8\textwidth]{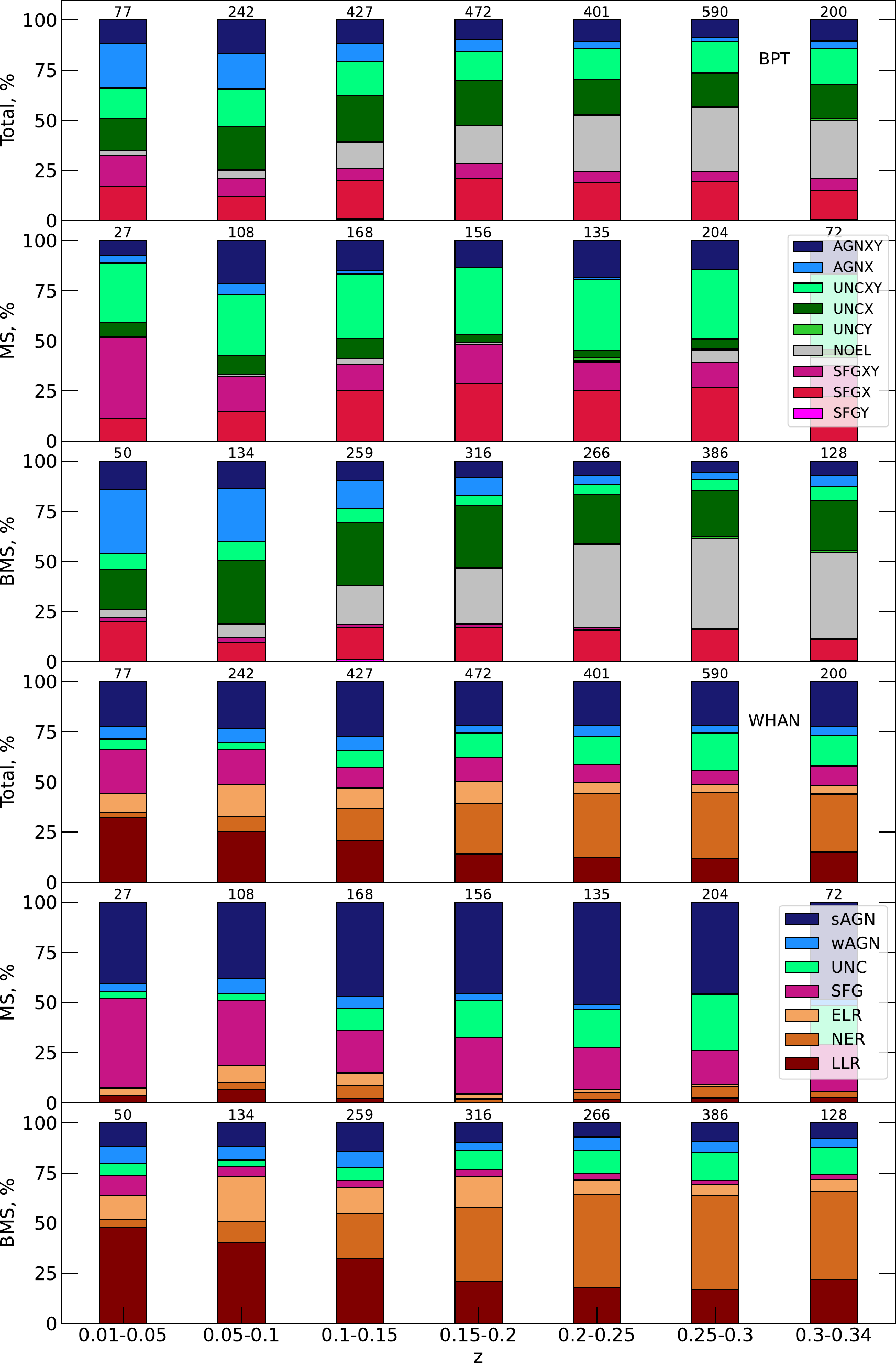}
\caption{BPT (top three panels) and the WHAN (bottom three panels) spectral class distribution as a function of redshift. In each group of panels, the top panel presents the distribution of the whole studied sample, the middle corresponds to the MS subsample, and the bottom displays the below-MS subsample. The information on colors can be found in legends.}
\label{fig:RED_BINS}
\end{figure*}

We started with the analysis of the distribution of the WHAN and BPT spectral classes in 6 redshift bins, presented in Fig.~\ref{fig:RED_BINS}. The structure of Fig.~\ref{fig:RED_BINS} is similar to Fig.~\ref{fig:AGE_BINS}: the top three panels display the BPT classifications for the whole studied sample, the MS subsample and the below-MS subsample (from top to bottom). The bottom three panels contain similar column charts but with the WHAN classification. 

The analysis of the whole sample indicated that BPT NOEL and WHAN NER (correspondent classes, see Appendix~\ref{ax:corr}) galaxies tend to increase their fraction for higher redshifts, while BPT AGNX and WHAN LLR (see Appendix~\ref{ax:corr}) decrease. That might be explained as distant galaxies being fainter, which leads to higher fractional errors of the flux and equivalent width of the ${\rm H\alpha}$ line, which in turn leads to the galaxy having a higher ${\rm EW_{H\alpha}}$ upper limit, above the LLR definition.

A closer look at the MS subsample also revealed 
the steady increase of WHAN UNC fraction for higher redshifts, which might also be explained by more distant galaxies possessing higher uncertainties for ${\mr EW_{\rm H\alpha}}$, hence putting an upper limit above the line that separates SFGs and AGNs from ELR at the WHAN diagram.

\begin{figure*}[h]
\centering
\includegraphics[width=0.95\textwidth]{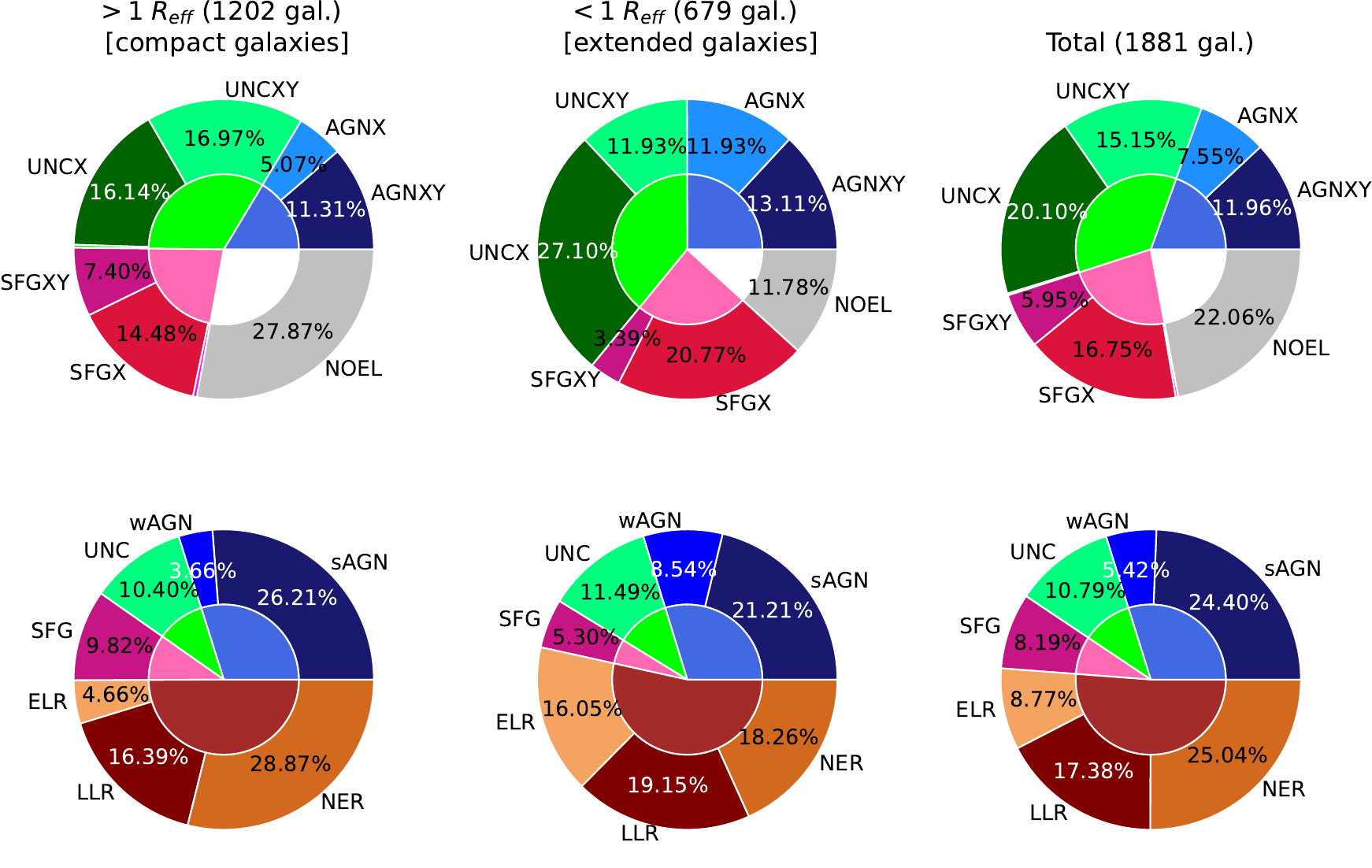}
\caption{The distribution of BPT (top panels) and WHAN (bottom panels) classes in the subsample with galaxies with spectrally observed part bigger than $1 R_{\rm eff}$ in r-filter (left), smaller than $1 R_{\rm eff}$ (middle), and for the total sample (right). Analogously to Fig. \ref{fig:BMSMS}, the outer circle displays detailed information on the spectral classes, whereas general classification into AGN, SFG, UNC, and NOEL/RGs is represented in the inner circle.
} 
\label{fig:AP_DIAG}
\end{figure*}

Next, recent work by \citet{Alban2023} inspired us to perform the analysis of WHAN and BPT fractions for subsamples of galaxies in which the fiber covers less than $1 R_{\rm eff}$ \citep[effective radius/half-light radius;][]{Vaucouleurs1978} in $r$-filter (extended galaxies, $<1 R_{\rm eff}$), and beyond (compact galaxies, $>1 R_{\rm eff}$). Galaxies with either a lack of the measurement of their effective radius or uncertainty of the measurement of $>50\%$
were excluded from the analysis. That was the case for $528$ early-type galaxies out of $2409$. The right panels of Fig.~\ref{fig:AP_DIAG} are similar to those of Fig.~\ref{fig:BMSMS}, thus excluding these galaxies does not alter the overall statistical results, and does not introduce a statistical bias.


The top row of Fig.~\ref{fig:AP_DIAG} contains information on the BPT classes in compact (left) and extended (middle) galaxies subsamples. Although the fraction of AGNXY galaxies is similar for both compact and extended galaxies ($\sim 12\%$), other BPT classes show significant differences between subsamples. Extended galaxies have higher AGNX, UNCX, and SFX fractions, while UNCXY, SFXY, and NOEL dominate in compact galaxies.

The distribution of WHAN spectral classes also suggests aperture effects. Compact galaxies are mostly classified as NER, sAGNs, and SFGs, while the extended galaxies subsample possesses higher ELR and wAGN fractions.

The median value of redshift for the compact galaxies subsample is $z_{\rm comp} = 0.216^{+0.073}_{-0.087}$, while for the extended galaxies it is $z_{\rm ext} = 0.147^{+0.104}_{-0.073}$, hence, compact galaxies are more distant than extended ones. That explains the higher BPT NOEL and WHAN NER fractions in compact galaxies, and oppositely, higher BPT X-objects and WHAN ELR fractions in extended galaxies subsample (see the discussion of the spectral class distribution within redshift bins above). 

The physical size might be the main parameter affecting the distributions of classes in compact and extended galaxy subsamples. However, assuming constant surface density, compact galaxies should be fainter and less massive than extended ones, 
but this is not the case, as the median values of stellar masses are similar: $M_{\rm s, \, comp} = 10.70^{+0.31}_{-0.43}$ and $M_{\rm s, \, ext} = 10.85^{+0.29}_{-0.35}$.

To sum up, the aperture analysis suggests that less distant galaxies with better resolved inner regions have a lower fraction of sAGNs and SFGs, while the share of ELR and wAGN are higher. This might be explained by the dilution of sAGN and SFG emission for apertures that include bigger parts of galaxies. Taking into account the conclusion of \citet{Alban2023}, the fraction of wrongly classified AGNs via the BPT diagram should increase significantly for apertures bigger than $1 \: R_{\rm eff}$ (compact galaxies), due to the high diffused ionized gas contamination. Hence, the fraction of SFGs and AGNs in our sample is overestimated only by several percent, which is an additional argument for the non-dominant role of AGN feedback and recent star-forming processes in our sample.

\section{Heavy dust attenuation}
\label{ax:dust}

\begin{figure*}[h]
\centering
\includegraphics[width=\textwidth]{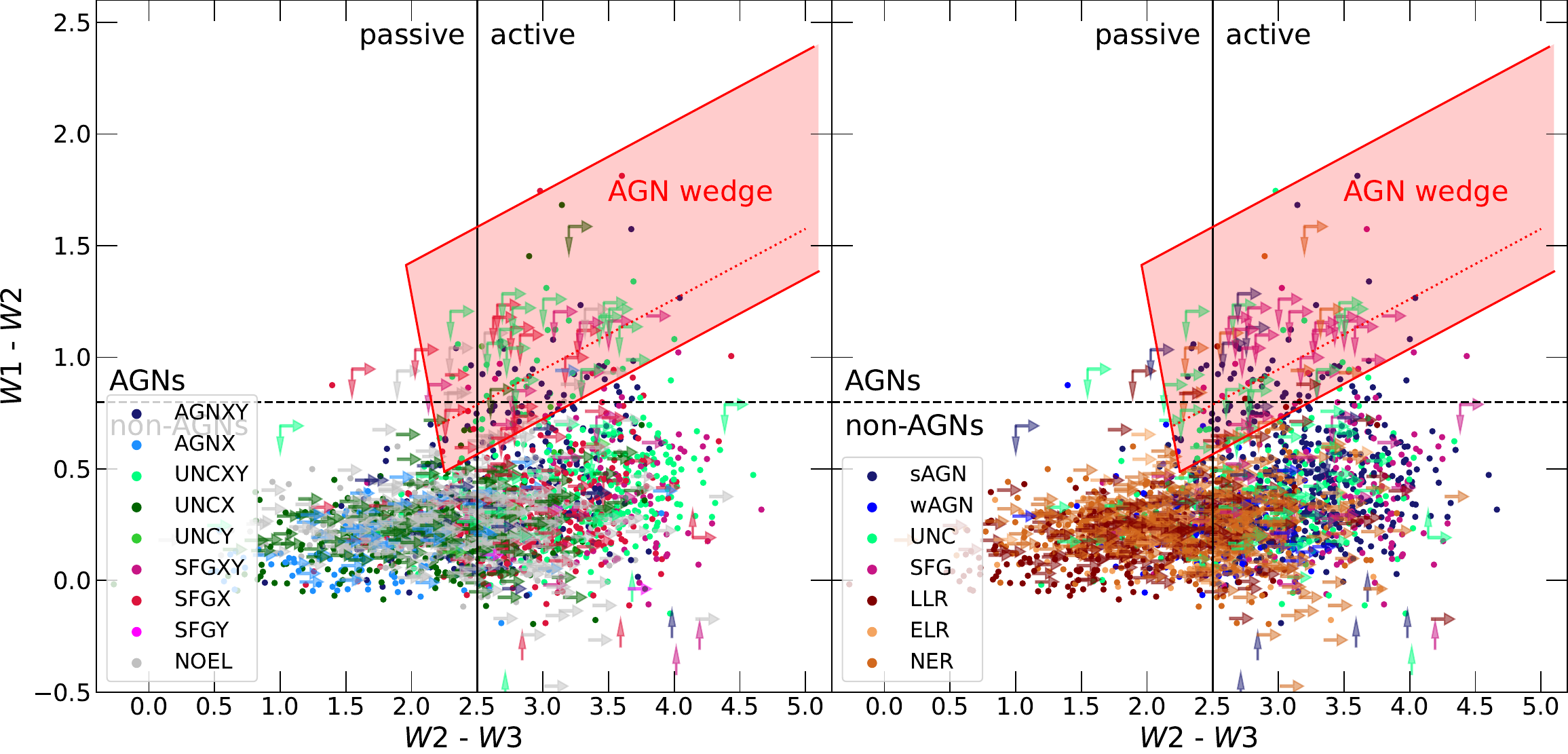}
\caption{The W1-W2 versus W2-W3 WISE color-color diagram with galaxies from the studied sample, color-coded by their BPT (left) and WHAN (right) spectral classes. The black solid line represents the division between star-forming and quenched galaxies, defined by \citet{Herpich_2016}. The black dashed line represents the lower limit for AGNs in WISE colors \citep{Stern2012}. The red polygon displays the so-called 'AGN wedge' \citep{Mateos2012}, with a red dotted line, determining the MIR power-law locus.} 
\label{fig:WISE}
\end{figure*}

Both BPT and WHAN diagnostics are immune to dust attenuation because lines used in ratios have similar wavelengths. The same is true for ${\mr EW_{\rm H\alpha}}$, for which the continuum and line fluxes are reduced by the same value due to the dust influence. However, heavy dust obscuration of AGN might result in a lack of detected emission lines in the spectra, which leads to misclassification of AGN \citep{Agostino2019}. 

In order to investigate the possible effect of this caveat, we make use of the WISE three-band classification. Fig.~\ref{fig:WISE} displays the galaxies from the studied sample placed in this diagram, color-coded by their BPT (left) and WHAN (right) classes. 

The AGN wedge \citep{Mateos2012} contains 39 BPT AGNXY and 58 WHAN sAGNs, as well as 35 BPT non-AGN and 16 WHAN non-AGN galaxies. The other AGN delimitation line \citep[black dashed line;][]{Stern2012} suggests that there are 26 BPT AGNXY and 41 WHAN sAGNs above it, as well as 31 BPT non-AGNs and 16 WHAN non-AGNs. That gives 26 and 32 WISE AGNs applying the \citealt{Stern2012} and \citealt{Mateos2012} thresholds, respectively; that are classified as non-AGN both via BPT and via WHAN. 

Conversely, the WHAN and BPT-defined AGNs out of the AGN area in the diagram might be explained with not enough dust content to detect the traces of its heating by AGN in WISE colors. This is confirmed by Fig.~\ref{fig:MDMS}, where both WHAN AGNs and BPT AGNXY tend to have lower dust mass fractions than other classes.  

To conclude, the WISE color diagnostics analysis suggested that the number of misclassified AGNs by the BPT and WHAN diagrams due to the heavy dust attenuation is negligible in our sample.

\section{Broad-line region emission}

\begin{figure*}[h]
\centering
\includegraphics[width=0.8\textwidth]{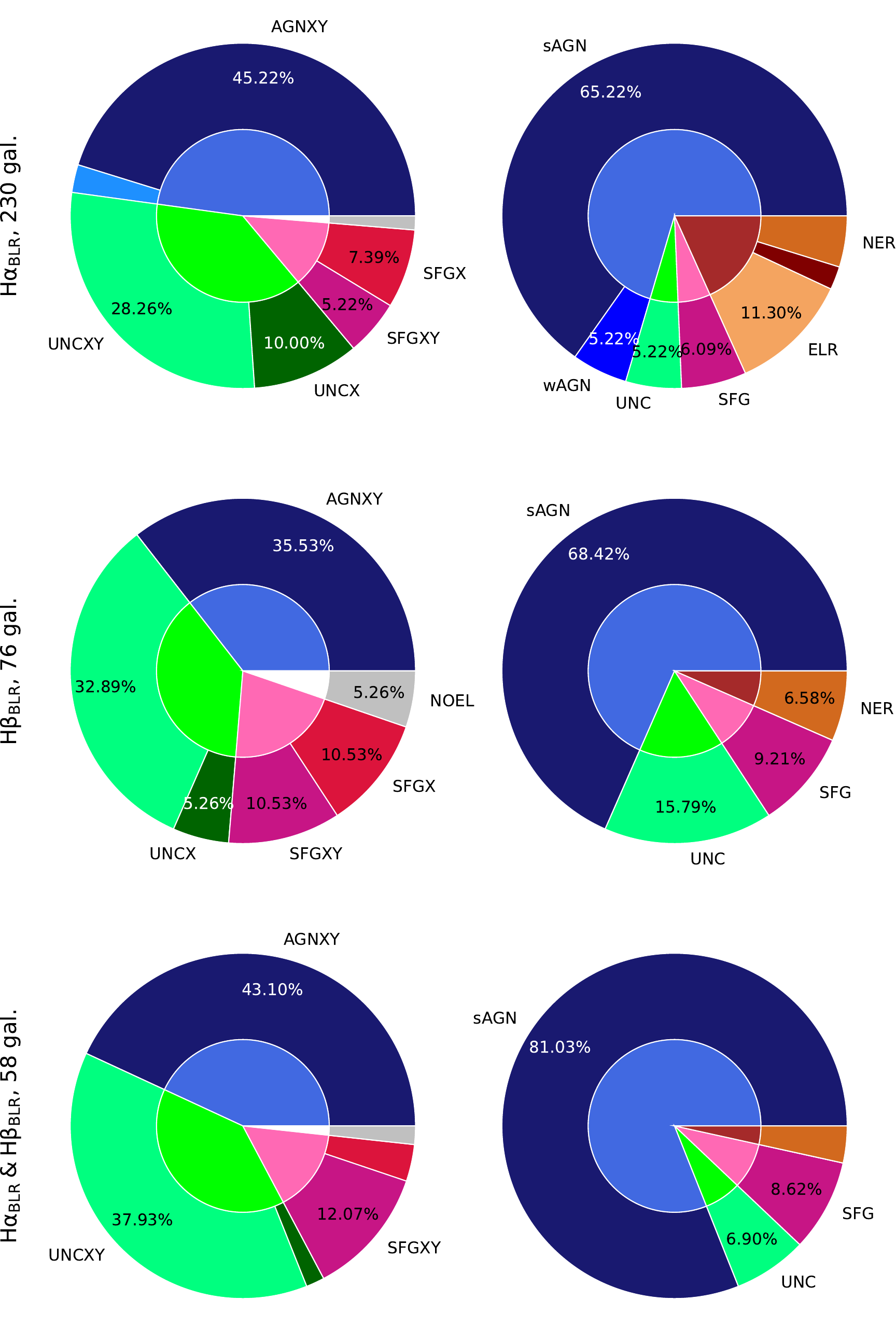}
\caption{The distribution of BPT (left) and WHAN (right) classes in the subsample of galaxies with broad plus narrow emission in ${\rm H\alpha}$ (230 galaxies, top row), in ${\rm H\beta}$ (76 galaxies, middle row) and in both ${\rm H\alpha}$ and ${\rm H\beta}$ simultaneously (58 galaxies, bottom row).}
\label{fig:BLR}
\end{figure*}

Finally, misclassification of AGN via emission line ratios might happen due to the broadening of the Balmer series due to the BLR emission, shifting a galaxy to the star-forming region at both BPT and WHAN diagrams. Although \citet{Agostino2019} showed the negligible effect of such a phenomenon, we decided to analyze it for the studied sample.

GAMA team performed simple and complex Gaussian fits\footnote{\url{https://www.gama-survey.org/dr3/data/cat/SpecLineSFR}} to several emission lines, including ${\rm H\alpha}$ and ${\rm H\beta}$. Based on their measurements, we selected the galaxies from the studied sample by the following criteria: 1) the model score of broad plus narrow emission versus single narrow emission should be positive; 2) the model score of continuum plus line fit versus only continuum should be positive; 3) fluxes and widths of a broad component should have uncertainties lower than $50\%$. Galaxies that had markers indicating problems with fitting narrow and broad components were also excluded from the sample. All the galaxies have the estimation of velocity dispersion of the broad component in the range between $500-5000 \: {\rm km/s}$, which indicates their origin from the BLR region of an AGN \citep{Peterson2006}. 

Figure~\ref{fig:BLR} displays the distribution of BPT (left) and WHAN (right) classes for 230 galaxies with detected broad and narrow emission in ${\rm H\alpha}$, 76 galaxies with broad components observed in ${\rm H\beta}$ and 58 galaxies that have broad components in both ${\rm H\alpha}$ and ${\rm H\beta}$. 

The BPT classification shows that $35-45\%$ of each broad-lined subsample is classified as AGNXYs, which correlates with the nature of emission, however, UNC galaxies and SFGs account for $30-40\%$ and $10-20\%$, respectively. On the other hand, $70-80\%$ of broad-lined subsamples are WHAN sAGNs and wAGNs, leaving $20-30\%$ for UNC, SFG, and even several RGs.

Thus, the misclassification of AGN due to the BLR emission is observable for the studied sample and increases the number of AGNs by $\sim4\%$ for BPT ($\sim100$ galaxies), and for WHAN by $\sim2\%$ ($\sim50$ galaxies).


\section{Additional figures and tables}

We present tables, containing a list of abbreviations used in the article (Table~\ref{tabABBR}),  distribution of BPT (Table~\ref{tabBPT}) and WHAN (Table~\ref{tabWHAN}) classes within age bins (Section~\ref{res:agesnap}) for the studied sample. We also present the 5 example rows from the machine-readable table, containing spectral classifications and outflow estimations for all $\sim300\,000$ GAMA galaxies with spectroscopic observations provided (Table~\ref{tabMRT}).

\begin{table*}[h]
\caption{Definitions of abbreviations, used in the manuscript.}
\begin{center}
    \begin{tabular}{c|l}
    \hline
    \hline
    Abbreviation & Definition \\
    \hline
    \multicolumn{2}{c}{Basic terms} \\
    \hline
        AGN & Active galactic nucleus \\
        BPT & Baldwin-Phillips-Telrevich diagram/classification, \citep{bpt}\\
        ETG & Early-type galaxy\\
        GAMA & Galaxy And Mass Assembly survey \\
        HOLMES & Hot low-mass evolved stars\\
        MRT & Machine-readable table\\
        MS & Main sequence, adopted from \citet{Speagle_2014}\\
        ISM & Interstellar medium \\
        PNe & Planetary nebulae\\
        RG & Retired galaxy\\
        S/N & Signal-to-noise \\
        SNe & Supernovae \\
        SFG & Star-forming galaxy \\
        SFR & Star-formation rate \\
        WHAN & Width of $\rm{H\alpha}$ vs. \NNf/$H\alpha$ diagram/classification, \citep{Cid_Fernandes_2011}\\
        WISE & Wide-field Infrared Survey Explorer \\
    \hline
    \multicolumn{2}{c}{BPT spectral classes} \\
    \hline
        AGNXY & Galaxies, classified as AGN hosts, using \NNf$/\rm{H\alpha}$ and \OOf$/\rm{H\beta}$ \\
        AGNX & Galaxies, classified as AGN hosts, using only \NNf$/\rm{H\alpha}$, and lacking detections of \OO and $\rm{H\beta}$\\
        AGNY & Galaxies, classified as AGN hosts, using only \OOf$/\rm{H\beta}$, and lacking detections of \NN and $\rm{H\alpha}$\\
        SFGXY & Galaxies, classified as SFGs, using \NNf$/\rm{H\alpha}$ and \OOf$/\rm{H\beta}$ \\
        SFGX & Galaxies, classified as SFGs, using only \NNf$/\rm{H\alpha}$, and lacking detections of \OO and $\rm{H\beta}$\\
        SFGY & Galaxies, classified as SFGs, using only \OOf$/\rm{H\beta}$, and lacking detections of \NN and $\rm{H\alpha}$\\
        UNCXY & Galaxies possessing at least one line detection in both \NNf$/\rm{H\alpha}$ and \OOf$/\rm{H\beta}$, \\
        & but these ratios allow both AGN and SFG classification\\
        UNCX & The \NNf$/\rm{H\alpha}$ ratio of these galaxies allow both AGN and SFG classification, lack of \OO and $\rm{H\beta}$ detection\\
        UNCY & The \OOf$/\rm{H\beta}$ ratio of these galaxies allow both AGN and SFG classification, lack of \NN and $\rm{H\alpha}$ detection\\
        NOEL & No emission among \OOf, $\rm{H\beta}$, \NN and $\rm{H\alpha}$ detected \\
    \hline
    \multicolumn{2}{c}{WHAN spectral classes} \\
    \hline
        sAGN & Host of strong AGN\\
        wAGN & Host of weak AGN\\
        SFG & Star-forming galaxy\\
        UNC & Unclear cases, where data allows both SFG and AGN, or both ELR and SFG/AGN classification\\
        ELR & Emission-line retired\\
        LLR & Line-less retired\\
        NLR & Noisy-line retired. Cases, where $\rm{EW_{H\alpha}}$ allows both ELR and LLR classification\\
        \hline
    \end{tabular}
\label{tabABBR}
\end{center}
\end{table*}

\begin{table*}[h]
\caption{The distribution of the BPT spectral classes within the age bins, described in Section~\ref{res:agesnap}.}
\begin{center}
\begin{tabular}{c|cc|cc|cc|cc|cc|cc|cc|cc}
\hline\hline
BPT class & \multicolumn{2}{c|}{$<8.8$} & \multicolumn{2}{c|}{8.8 -- 9.0} & \multicolumn{2}{c|}{9.0 -- 9.2} & \multicolumn{2}{c|}{9.2 -- 9.4} & \multicolumn{2}{c|}{9.4 -- 9.6} & \multicolumn{2}{c|}{9.6 -- 9.8} & \multicolumn{2}{c|}{9.8 -- 10.0} & \multicolumn{2}{c}{Tot. in class}\\
 & \multicolumn{2}{c|}{dex(yr)} & \multicolumn{2}{c|}{dex(yr)} & \multicolumn{2}{c|}{dex(yr)} & \multicolumn{2}{c|}{dex(yr)} & \multicolumn{2}{c|}{dex(yr)} & \multicolumn{2}{c|}{dex(yr)} & \multicolumn{2}{c|}{dex(yr)} & \\ 
\hline
AGNXY & 4 & & 15 & & 40 & & 70 & & 89 & & 38 & & 6 & & 262 & \\ 
AGNX & 0 & \textbf{4} & 1 & \textbf{16} & 4 & \textbf{44} & 13 & \textbf{83} & 36 & \textbf{125} & 81 & \textbf{119} & 25 & \textbf{31} & 160 & \textbf{422}\\ 
AGNY & 0 & & 0 & & 0 & & 0 & & 0 & & 0 & & 0 & & 0 & \\ 
\hline

UNCXY & 21 & & 49 & & 92 & & 128 & & 73 & & 23 & & 0 & & 386 & \\ 
UNCX & 0 & \textbf{22} & 4 & \textbf{54} & 7 & \textbf{102} & 55 & \textbf{184} & 182 & \textbf{260} & 200 & \textbf{223} & 22 & \textbf{22} & 470 & \textbf{867}\\ 
UNCY & 1 & & 1 & & 3 & & 1 & & 5 & & 0 & & 0 & & 11 & \\ 
\hline

SFGXY & 26 & & 34 & & 37 & & 38 & & 17 & & 5 & & 0 & & 157 & \\ 
SFGX & 10 & \textbf{36} & 12 & \textbf{46} & 49 & \textbf{86} & 143 & \textbf{181} & 187 & \textbf{209} & 36 & \textbf{41} & 7 & \textbf{7} & 358 & \textbf{606}\\ 
SFGY & 0 & & 0 & & 0 & & 0 & & 5 & & 0 & & 0 & & 5 &\\ 
\hline

NOEL & \multicolumn{2}{c|}{\textbf{3}} & \multicolumn{2}{c|}{\textbf{4}} & \multicolumn{2}{c|}{\textbf{9}} & \multicolumn{2}{c|}{\textbf{69}} & \multicolumn{2}{c|}{\textbf{206}} & \multicolumn{2}{c|}{\textbf{212}} & \multicolumn{2}{c|}{\textbf{11}} & \multicolumn{2}{c}{\textbf{514}}\\
\hline

Tot. in bin & \multicolumn{2}{c|}{\textbf{65}} & \multicolumn{2}{c|}{\textbf{120}} & \multicolumn{2}{c|}{\textbf{241}} & \multicolumn{2}{c|}{\textbf{517}} & \multicolumn{2}{c|}{\textbf{800}} & \multicolumn{2}{c|}{\textbf{595}} & \multicolumn{2}{c|}{\textbf{71}} & \multicolumn{2}{c}{\textbf{2409}}\\

\hline
\end{tabular}
\label{tabBPT}
\end{center}
\end{table*}

\begin{deluxetable*}{c|ccccccc|c}
\label{tabWHAN}
\tabletypesize{\scriptsize}
\tablecaption{The distribution of the WHAN spectral classes within the age bins, described in Section~\ref{res:agesnap}.}
\tablenum{3}
\tablehead{
\colhead{WHAN class} & \colhead{$<$8.8} & \colhead{8.8 -- 9.0} & \colhead{9.0 -- 9.2} & \colhead{9.2 -- 9.4} & \colhead{9.4 -- 9.6} & \colhead{9.6 -- 9.8} & \colhead{9.8 -- 10.0} & \colhead{Tot. in class}\\
\colhead{} & \colhead{dex(yr)} & \colhead{dex(yr)} & \colhead{dex(yr)} & \colhead{dex(yr)} & \colhead{dex(yr)} & \colhead{dex(yr)} & \colhead{dex(yr)} & \colhead{}}
\startdata
sAGN & $14$ & $42$ & $104$ & $198$ & $157$ & $35$ & $3$ & 553  \\
wAGN & $1$ & $0$ & $5$ & $19$ & $72$ & $23$ & $3$ & 123  \\
UNC & $5$ & $14$ & $43$ & $104$ & $112$ & $23$ & $3$ & 304  \\
SFG & $41$ & $56$ & $64$ & $63$ & $29$ & $4$ & $0$ & 257  \\
ELR & $0$ & $1$ & $3$ & $16$ & $82$ & $82$ & $10$ & 194  \\
NER & $3$ & $4$ & $14$ & $78$ & $242$ & $235$ & $14$ & 590  \\
LLR & $1$ & $3$ & $8$ & $39$ & $106$ & $193$ & $38$ & 388  \\
\hline
Tot. in bin & $65$ & $120$ & $241$ & $517$ & $800$ & $595$ & $71$ & 2409  \\
\enddata
\end{deluxetable*}

\begin{splitdeluxetable*}{cccccccccBcccccc}

\tabletypesize{\scriptsize}
\tablecaption{The five data rows from the machine-readable table (MRT) are presented. The columns, from right to left: unique GAMA ID, RA, DEC, spectroscopic redshift, distance from the MS, BPT, and WHAN spectral class, and percentiles of the outflow estimation with AGN luminosity considered (`AGN on') and not (`AGN off'). MRT contains the estimation of listed parameters for all galaxies with performed spectral observations of them (305\,529 galaxies).}

\tablenum{4}

\tablehead{\colhead{CATAID} & \colhead{RA} & \colhead{DEC } & \colhead{z} & \colhead{${\rm \Delta MS}$} & \colhead{BPT } & \colhead{WHAN } & \colhead{$n$ } & \colhead{${ n_{\rm er}}$ } & \colhead{out\_on\_16 } & \colhead{out\_on\_50} & \colhead{out\_on\_84} & \colhead{out\_off\_16} & \colhead{out\_off\_50} & \colhead{out\_off\_84} \\ 
\colhead{} & \colhead{(deg)} & \colhead{(deg)} & \colhead{} & \colhead{([${\rm M_\odot/yr}$])} & \colhead{} & \colhead{} & \colhead{} & \colhead{} & \colhead{([${\rm M_\odot/yr}$])} & \colhead{([${\rm M_\odot/yr}$])} & \colhead{([${\rm M_\odot/yr}$])} & \colhead{([${\rm M_\odot/yr}$])} & \colhead{([${\rm M_\odot/yr}$])} & \colhead{([${\rm M_\odot/yr}$])} } 
\startdata
491414 & 211.50587 & -1.53749 & 0.07012284 & -1.934 & NOEL & LLR & 4.2187 & 0.1227 & -100000.0 & -0.099 & -100000.0 & -2.873 & -2.12 & -1.662 \\
484604 & 211.41319 & -1.89285 & 0.03538566 & -99999.0 & SFGXY & SFG & -99999.0 & -99999.0 & -99999.0 & -99999.0 & -99999.0 & -99999.0 & -99999.0 & -99999.0 \\
484621 & 211.63321 & -1.96885 & 0.0350057 & -0.29 & AGNXY & ELR & 2.4309 & 0.0144 & 0.134 & 0.235 & 0.319 & -0.472 & -0.31 & -0.156 \\
463148 & 211.68703 & -1.33978 & 0.0534042 & -0.153 & AGNXY & sAGN & 2.3842 & 0.0339 & 1.089 & 1.122 & 1.152 & -0.319 & -0.173 & -0.097 \\
735968 & 211.59354 & -1.20857 & 1.1536546 & 0.288 & NDA & NDA & 3.7668 & 0.7839 & -99999.0 & -99999.0 & -99999.0 & 2.306 & 2.306 & 2.306 \\
\enddata


\tablecomments{$-99999.0$ flag indicates the lack of the data to calculate a parameter. $-100000.0$ in 16/84th outflow percentile means that the 50th percentile of the outflow estimation is presented as an $2\sigma$ upper limit. The NDA flag in the BPT and WHAN columns stands for problems in the estimation of the flux of lines, involved in the correspondent classification. The Table is published in its entirety in the machine-readable format. A portion is shown here for guidance regarding its form and content.}

\label{tabMRT}
\end{splitdeluxetable*}

\end{document}